\documentclass[12pt]{article}

\usepackage{amsmath, setspace}
\usepackage{amssymb, xspace}

\usepackage{graphicx}

\usepackage{color} 

\usepackage{hyperref}
\usepackage{times}
\usepackage[round]{natbib}
\usepackage[table,rgb]{xcolor}

\usepackage[labelfont=bf,labelsep=period,justification=raggedright]{caption}

\makeatletter
\renewcommand{\@biblabel}[1]{\quad#1.}
\makeatother

\date{}

\newcommand{\fg}{Fig.~\ref}

\definecolor{citecolor}{rgb}{0.05, 0.05, 0.5}
\definecolor{linkcolor}{rgb}{0.2, 0.05, 0.05}
\hypersetup{
colorlinks=true, 
citecolor=citecolor,
linkcolor=linkcolor
}
\graphicspath{ {figs/}
}

\begin{document}

\title{Extensive Regulation of Metabolism and Growth during the Cell Division Cycle}
\title{Extensive regulation of metabolism and growth during the cell division cycle}
\maketitle


\begin{flushleft}
{\large
Nikolai Slavov$^{1,2,\dagger}$,
David Botstein$^{2}$ and
Amy Caudy$^{2,3}$
} 

{\footnotesize 
$1$ Departments of Physics and Biology, Massachusetts Institute of Technology, Cambridge, MA 02139, USA \\
$2$ Lewis-Sigler Institute for Integrative Genomics and Molecular Biology Department, Princeton University, Princeton, NJ 08544, USA \\
$3$ Donnelly Centre for Cellular and Biomolecular Research, University of Toronto, Toronto, M5S 3E1, Canada \\
$\dagger$ Corresponding author: nslavov@alum.mit.edu
}
\end{flushleft}
{\bf Abbreviations:} CDC - cell division cycle; GR - growth rate; HOC - high oxygen consumption; LOC - low oxygen consumption; EAP - ensemble average over phases; 
\\



\newpage 

\section*{Abstract}
Yeast cells grown in culture can spontaneously synchronize their respiration, metabolism, gene expression and cell division. Such metabolic oscillations in synchronized cultures reflect single-cell oscillations, but the relationship between the oscillations in single cells and synchronized cultures is poorly understood. To understand this relationship and the coordination between metabolism and cell division, we collected and analyzed DNA-content, gene-expression and physiological data, at hundreds of time-points, from cultures metabolically-synchronized at different growth rates, carbon sources and biomass densities. The data enabled us to extend and generalize an ensemble-average-over-phases (EAP) model that connects the population-average gene-expression of asynchronous cultures to the gene-expression dynamics in the single-cells comprising the cultures. The extended model explains the carbon-source specific growth-rate responses of hundreds of genes. Our data demonstrate that for a given growth rate, the frequency of metabolic cycling in synchronized cultures increases with the biomass density. This observation underscores the difference between metabolic cycling in synchronized cultures and in single cells and suggests entraining of the single-cell cycle by a quorum-sensing mechanism. Constant levels of residual glucose during the metabolic cycling of synchronized cultures indicate that storage carbohydrates are required to fuel not only the G1/S transition of the division cycle but also the metabolic cycle. Despite the large variation in profiled conditions and in the time-scale of their dynamics, most genes preserve invariant dynamics of coordination with each other and with the rate of oxygen consumption. Similarly, the G1/S transition always occurs at the beginning, middle or end of the high oxygen consumption phases, analogous to observations in human and drosophila cells. These results highlight evolutionary conserved coordination among metabolism, cell growth and division.

\begin{spacing}{1.4}
\newpage

\section*{Introduction}
The coordination of cell growth with cell division \citep{polymenis1997coupling, neufeld1998connections, jorgensen2004cells, fingar2004target, tzur2009cell, goranov2010growth, brauer_2008, Slavov_eth_grr, hoose2012systematic, Jan_cell_size, Slavov_aux} and the deregulation of these processes in cancer \citep{cairns_regulation_2011, dang2012metabolism_and_cancer, Jain25052012} have been studied since the nineteenth century \citep{wilson1900cell}. Despite major insights, such as the dependence of cell division on cell growth \citep{johnston1977coordination, brauer_2008, Slavov_eth_grr} and the identification of participating pathways \citep{jorgensen2002systematic,Polymenis-division-control}, many aspects of the coordination between cell growth and division remain unclear \citep{jorgensen2004cells,benanti2012coordination, Slavov_exp}. 
    To investigate this coordination, we exploited the spontaneous metabolic synchronization of yeast cultures: when a glucose-starved culture of budding yeast is refed with a glucose-limited medium at a constant rate, the culture can begin a respiratory cycle, manifested by periodic oscillation in the levels of dissolved oxygen in the culture medium \citep{ymc_1969,Kaspar_1969} and periodic expression of thousands of genes \citep{klevecz_genomewide_2004, tu_logic_2005, murray2007regulation, Slavov_batch_ymc, Murray2012yin}, reviewed in \citep{tu2006_review, murray_tuneable_2007, Laxman2010, murray_redox_2011}. Recently we observed similar oscillations in phosphate-starved nondividing yeast cells \citep{Slavov_batch_ymc}, demonstrating that these phenomena of metabolic synchronization are not restricted to carbon-source-limited or continuous yeast cultures. Rather, multiple lines of evidence including ($i$) single-cell analysis \citep{ymc_2010, wyart_evaluating_2010}, ($ii$) a mechanistic model connecting the growth-rate and stress responses of synchronized and asynchronous cultures \citep{brauer_2008, Slavov_eth_grr, Slavov_batch_ymc, Slavov_emc}, and ($iii$) elutriation-synchronized cultures \citep{Slavov_emc} suggest that the cycling in metabolically synchronized cultures reflects a metabolic cycle in single cells from asynchronous cultures.

This metabolic cycling in single cells, however, does not exactly parallel the metabolic cycling in synchronized cultures since measurements of the DNA content of single cells from the synchronized cultures \citep{klevecz_genomewide_2004, tu_logic_2005, robertson_real-time_2008, Polymenis-mtDNA-sulfur, Slavov_eth_grr, genome-wide-transcriptional-oscillators} have demonstrated at least two sub-populations during each period of the synchronized cultures: dividing (2C DNA content) and non-dividing (1C DNA content). Thus, distinguishing between metabolic cycling in synchronized cultures and in single cells is crucial to the analysis throughout this article. Early studies in elutriation-synchronized cultures suggest the metabolic cycling in single cells includes cycling of biosynthetic enzymes, mRNAs for most biosynthetic genes and the translation rate \citep{creanor1982patterns,mitchison1969enzyme,Slavov_emc}. 

The dividing and the nondividing subpopulations that coexist during each metabolic cycle in synchronized cultures demand efforts to identify the genes and processes unique to either growth or division. To identify such genes and clarify the difference between the metabolic cycling in single cells and in synchronized cultures, we exploited our earlier observation that the phases of distinct subpopulations, i.e., dividing and nondividing, shift between different growth conditions \citep{Slavov_eth_grr}. We metabolically synchronized cultures of budding yeast over a wide range of growth conditions, differing by their growth rates and biomass densities, and characterized their gene expression and physiology. We found that the frequency of cycling in metabolically synchronized cultures increases with biomass density, reinforcing the difference between metabolic cycling in single cells and in synchronized cultures and providing the first evidence for quorum sensing in budding yeast whose growth is limited by the carbon source (glucose) and not by nitrogen. Using these findings and our gene expression data, we developed a general model connecting cell division and metabolic cycling in single cells to the entrained metabolic cycles observed in synchronized cultures. This model can account for the measured gene expression over hundreds of time points in both metabolically synchronized and in asynchronous cultures growing at different growth rates, biomass densities, nutrient limitations, and carbon sources. 

\section*{Results}
\subsubsection*{Experimental Conditions} 
We metabolically synchronized yeast as described previously by \citet{ymc_1969, Slavov_eth_grr}: we grew the \emph{DBY12007} diploid strain in glucose-limited media to stationary phase and subsequently fed the cultures with glucose-limited media at constant rates. The experimental factors that we regulated to achieve a variety of growth conditions were ($i$) the concentration of glucose in the feed media, which determines the biomass density \citep{ferea1999systematic, Slavov_eth_grr, slavov_thesis}, and ($ii$) the dilution rate, which at steady-state (or the oscillatory [limit-cycle] regime) equals the average growth rate of the cultures. Under these conditions, our cultures synchronized metabolically, as manifested by the periodic oscillation in the levels of dissolved oxygen in the culture media, \fg{fig:CDC_YMC_periods}A. Such measurements of dissolved oxygen track closely with the oxygen consumption rate, as demonstrated by independent measurements of oxygen in the exhaust gas 
                 \citep{murray_redox_2011, 2012_Murray} and by the observation that injecting a carbon source into a carbon-source-limited culture induces a precipitous drop of the dissolved oxygen within seconds \citep{ronen_transcriptional_2006, slavov_thesis}. The trace of dissolved oxygen in these metabolically synchronized cultures can be divided phenomenologically into two phases, the low oxygen consumption phase (LOC), when the amount of oxygen in the media is high because the cells consume little oxygen, and the high oxygen consumption phase (HOC), when the reverse holds. This division can be quantified rigorously using the distribution of dissolved oxygen levels, \fg{fig:CDC_YMC_periods}B. The distribution is bimodal because a metabolically synchronized culture spends most of the time either in HOC (left mode in \fg{fig:CDC_YMC_periods}B) or in LOC (right mode \fg{fig:CDC_YMC_periods}B) and much less in the transition (the data points between the modes).

\subsubsection*{The Period of Metabolic Cycling in Synchronized Cultures Depends Linearly on the Growth Rate and on the Glucose Concentration in the Feed Media}
We sought to explore how the period of metabolic cycling of synchronized cultures depends on the growth rate (doubling period of the culture) and on the concentration of glucose in the feed media (determining the biomass density) while all other parameters were kept constant, \fg{fig:CDC_YMC_periods}C. The data in  \fg{fig:CDC_YMC_periods}C reinforce the linear dependence between the doubling periods of the cultures and the periods of the metabolic cycle that we reported previously  \citep{Slavov_eth_grr} and show a striking new dependence: for a given growth rate, the higher the concentration of glucose in the feed media ($[Glu]_{feed}$), and thus the biomass density, the shorter the period of the metabolic cycle.  
This dependence is quantified by the decreasing intercepts of the linear models fit to the data, \fg{fig:CDC_YMC_periods}C. Such continuous dependence between the biomass density and the cycling period provides a mechanistic link between the ``short'' and the ``long'' period metabolic cycling in synchronized cultures that have been treated as different phenomena. In our experimental setup, we are able to regulate continuously the frequency of metabolic cycling of synchronized diploid cultures simply by changing the growth rate (dilution rate) or the biomass density ($[Glu]_{feed}$).  A haploid culture, however, failed to synchronize for all combinations of growth rates and biomass densities; see Discussion. The linear models in \fg{fig:CDC_YMC_periods}C also highlight a dependence between $[Glu]_{feed}$ and the rate at which the frequency of the metabolic cycling in synchronized cultures changes with the growth rate: the rate of change increases with increase of $[Glu]_{feed}$, as quantified by the increasing slopes. 

\subsubsection*{The Metabolic Cycling in Synchronized Cultures is Emergent Behaviour}
We observe that the frequency of metabolic cycling in synchronized cultures can be altered by changing the $[Glu]_{feed}$ while the growth rate ($\mu$) and the doubling period are held constant. The dependence of the frequency on $[Glu]_{feed}$, which determines the number of cells per ml in the culture \citep{ferea1999systematic, Slavov_eth_grr, slavov_thesis}, suggests that some form of cell-cell communication, dependent on population density, affects the population dynamics. Therefore, the dynamics in a synchronized culture do not map one-to-one to the dynamics in single cells but depend on cell-cell communication, and thus on the collective properties of the culture. Remarkably, the increase of the cell density increases the frequency of the metabolic cycling in the synchronized culture,  in full accord with the predictions of models based on quorum-sensing synchronization \citep{Strogatz_pnas_2004, schwab_dynamical_2010}. Examples of quorum-sensing have been described in the distantly related  \emph{Candida albicans} \citep{hornby2001quorum,2004_quorum_fink_Candida_albicans} and in \emph{Saccharomyces cerevisiae} whose growth is limited by nitrogen \citep{2006_quorum_fink}. Our data suggest that glucose-limited \emph{Saccharomyces cerevisiae} may also have a quorum-sensing mechanism. Different authors have shown that many different chemicals can affect the transitions between the phases of metabolically synchronized cultures \citep{murray1999glutathione, keulers1996synchronization_CO2, murray2003acetaldehyde_sulfite, Kuriyama2003_H2S, robertson_real-time_2008, tu2009evidence_CO2}; we do not know which ones of these chemicals, if any, mediate the quorum sensing that we observe.   

We find that at least two parameters can affect the frequency of metabolic cycling in synchronized cultures: the biomass density and the dilution rate, which in the oscillatory (limit-cycle) regime that we study equals the growth rate ($\mu$) of the culture. These two parameters, however, have very different effects on the duration of LOC relative to HOC. The increase in growth rate shortens  the duration of LOCs relative to HOC \citep{Slavov_eth_grr}. In contrast, increase in the biomass density shortens both LOC and HOC preserving their relative duration unchanged.  %

\subsubsection*{The Metabolic Cycling in Synchronized Cultures is Fueled by Reserved Carbohydrates and Depends on Cell Density}
The rate of glucose catabolism is higher during HOC than during LOC, since HOC is associated with higher rates of oxidative phosphorylation, as revealed by higher oxygen consumption, and also with higher rates of fermentation, as revealed by glycolytic fermentation \citep{Kaspar_1969}. On the other hand, a continuous metabolically synchronized culture has a constant influx of glucose from the feed media. If the rate of glucose import equals the rate of glucose consumption, the  concentration of glucose in the culture media would oscillate, peaking during LOC and reaching a nadir at the end of HOC. However, if the rate of glucose import equals the maximum rate of import that yeast can achieve and is thus constant, the  concentration of glucose in the culture media would be constant too, and the glucose used during the HOC phase must come from intracellular stores. To distinguish between these two possibilities, we  measured the concentration of residual glucose in the culture medium of a synchronized culture fed with a medium containing $800 mg/L$ glucose and growing at $\mu=0.133 h^{-1}$. We found that the residual glucose concentration  is constant across the metabolically synchronized cultures (\fg{fig:glu}) and in good agreement with the corresponding concentration measured by \citet{boer_influence_2008} for an asynchronous culture growing at this range of growth rates in a glucose-limited medium.
The culture has a constant influx of glucose from the feed media, and we measured a constant concentration of the residual glucose. Thus, mass conservation requires that the glucose uptake by the culture be constant too. This result is consistent only with the second mechanism and suggests that yeast maximize the rate of glucose import and store glucose as reserved carbohydrates when it is not consumed. This inference is consistent with the increased levels of intracellular trehalose and glycogen measured during the LOC phase in yeast \citep{ymc_1969,trehalose_2010}, and with similar observation in mammalian cells \citep{rousset1979presence}. Hence, reserved carbohydrates are not just accumulated throughout the G1 phase and used during S/G2/M phases of the CDC but they are also used dynamically during the metabolic cycle. The constant uptake of glucose most likely reflects that the cells are taking up the limiting resource, glucose, as fast as possible given the concentration in the media and their import capacity, \fg{fig:glu}.   


\subsubsection*{Variation in the Fraction of Cells Dividing per Metabolic Cycle in Synchronized Cultures }
Having demonstrated that the cycling in synchronized cultures is an emergent behavior that depends on cell density, we next sought to understand how the cycling of a synchronized culture relates to the subpopulations comprising the culture. To this end, we measured dissolved oxygen and DNA content in three continuous, glucose-limited, metabolically synchronized cultures growing at different biomass densities and growth rates. Figure \ref{fig:dna_content} displays the data from these cultures, together with the previously published \citep{Slavov_batch_ymc} analogous data from a batch culture growing in a phosphate-limited medium with ethanol as a sole source of carbon and energy. Each of the continuous cultures was cycling steadily so that any two time-points with the same phase had the same cell-density; thus, the fraction of cells dividing per metabolic cycle equals the ratio of the period of metabolic cycling ($T_{MC}$) to the CDC period ($T_{CDC}$), $\frac{T_{CDC}}{T_{MC}}$, as displayed below the growth parameters on the top of \fg{fig:dna_content}. In the batch culture, this ratio equals zero for the sampled period since no cells passed the G1/S checkpoint \citep{Slavov_batch_ymc}. These measurements of the fraction of dividing cells per metabolic cycle in the synchronized cultures are consistent with the DNA content measurements and demonstrate that our experimental setup has succeeded in achieving a substantial variation between the dividing and nondividing subpopulations across different conditions of metabolic synchrony.

\subsubsection*{The Difference in Oxygen Consumption between LOC and HOC Decreases with Growth Rate}
The dissolved oxygen data in \fg{fig:dna_content} have been normalized so that for each culture zero dissolved oxygen corresponds to the lowest measured dissolved oxygen (highest consumption) and $1$ corresponds to the dissolved oxygen in the medium before inoculation (zero consumption).
 These data indicate a substantial variation between the relative dissolved oxygen levels during the LOC and the HOC phases. At one extreme, the nondividing batch culture has extremely low oxygen consumption during LOC. At the other extreme, the fastest growing continuous culture has very high oxygen consumption in LOC relative to HOC. Generally, as growth rate increases, so does the oxygen consumption during LOC relative to HOC. A possible explanation for this observation is that as the growth rate increases and the relative duration of the LOC phase decreases \citep{Slavov_eth_grr}, the intensity of the physiological processes taking place in LOC and requiring energy, such as macromolecular recycling and repair, also increases to compensate for the shorter duration of the LOC phase. 

\subsubsection*{The CDC Phases Shifts Relative to the Metabolic Phases in Synchronized Cultures }
Since we have markers for both the division cycle (DNA content) and for the metabolic cycle (dissolved oxygen), we can observe and characterize the two cycles across the studied growth conditions. In the continuous cultures, the increase in the fraction of cells replicating DNA (G1/S transition; \fg{fig:dna_content}) always begins during HOC albeit with variation:  at the very beginning of the HOC phase ($\mu=0.10h^{-1}$ and $[Glu] = 300mg/L$), midway through the HOC phase ($\mu=0.133h^{-1}$ and $[Glu] = 800mg/L$) or somewhere in between ($\mu=0.095h^{-1}$ and $[Glu] = 800mg/L$); This variation is much more pronounced for the fraction of cells with duplicated DNA that peaks either during LOC or during HOC. This variation is reinforced by the data for the largely nondividing batch culture \citep{Slavov_batch_ymc}, \fg{fig:dna_content}. 

When each synchronized culture is considered independently, genes expressed in different subpopulations, i.e., dividing and nondividing, would appear to be correlated with a specific phase offset. As evident from \fg{fig:dna_content}, however, these correlations between the two subpopulations will change as the offset between the metabolic and the division cycles changes in different conditions.  By considering the data for all conditions together, we can distinguish condition-specific correlations from the condition-invariant correlations that hold across all growth conditions. For example, the levels of genes expressed in dividing cells (S/G2/M) should always be correlated to each other in the same way regardless of growth condition, i.e., have condition-invariant correlation. In contrast, the  correlations between these genes expressed in dividing cells and the genes expressed in the nondividing subpopulation are expected to change with the growth conditions \citep{Slavov_eth_grr} as the phase offsets between different subpopulations shift. Furthermore, we can associate the sets of genes that have condition-invariant correlations with either the cell division or the cell growth cycle based on their correlation to either DNA content (cell division) or dissolved oxygen (metabolism and cell growth). 


\subsubsection*{Coordination of Gene Expression Dynamics during Cell Growth and Division}    
To identify such sets of correlation-invariant genes, we used DNA microarrays to measure gene expression (see Methods) in all three cultures from  \fg{fig:dna_content}. We combined and clustered these data (\fg{fig:transcripts}) together  with previously published  gene-expression data from a batch metabolically-synchronized culture growing in  a phosphate-limited ethanol medium  \citep{Slavov_batch_ymc} and from a continuous glucose-limited culture \citep{tu_logic_2005}. The most striking observation in the clustered data (\fg{fig:transcripts}) is that despite the differences in growth conditions, the vast majority of genes are expressed in the same phase relative to the trace of dissolved oxygen.

In stark contrast, the expression levels of a small cluster of genes (marked with a blue bar in \fg{fig:transcripts}) shift relative to the phases of most other genes and the phase of the oxygen consumption. These genes fall exactly within the above definitions of having condition-invariant correlations to each other and  condition-specific correlations to most of the other genes, we selected them for further characterization, \fg{fig:cdc_div}A.
 All of these genes are annotated to the G2 and M phases of the CDC. 
Across all studied cultures, these genes correlate with cell division but not with the oxygen consumption, \fg{fig:cdc_div}A. This result demonstrates that the phase offsets between oxygen consumption and DNA content from \fg{fig:dna_content} are also reflected in similar offsets between the phases of oxygen consumption and the expression of a coherent cluster of G2/M phase genes. 
  
Notably, those genes that correlate with cell division are only a fraction of those annotated to the cell cycle based on their periodic expression in mitotically synchronized cells (Gene Ontology term $GO:0007049$) growing in rich media \citep{spellman_1998}. Thus we sought to explore the expression pattern of the remaining genes annotated to the CDC particularly since we previously observed \citep{Slavov_batch_ymc} that the majority of the genes annotated to the CDC did not correlate to the cell division as measured by DNA content. To test the generality of our previous observation, we clustered the CDC-annotated genes that are expressed periodically in the nondividing batch culture,  \fg{fig:cdc_div}B. The expression levels of these genes correlate strongly with the dissolved oxygen in the media but not with the fraction of cells with replicated DNA, reinforcing and generalizing our observations from the batch culture. The periodic transcription of many genes in concert with the metabolic rather than the cell division cycle emphasizes the metabolic changes inherent in growing cells and makes a clear distinction between growth and division. The extent to which this transcriptional oscillation is translated into a protein oscillation depends on translational control similar to the G1 cyclin Cln3 \citep{polymenis1997coupling}. Thus we conclude that most gene annotated to the CDC are transcribed in concert with metabolism.

To explore  gene expression patterns more globally and quantitatively, we decomposed the gene expression data into singular vectors (\fg{fig:svd}). These vectors are low dimensional projections of expression patterns that account for most of the variance ($\sigma^2$) in the data,  \fg{fig:svd}. The first singular vector accounting for nearly half of the variance in gene expression is very similar for all cultures and correlates in a very similar way to the dissolved oxygen (top panels in \fg{fig:svd}). This result strongly supports the correlation between gene expression and oxygen consumption observed in \fg{fig:transcripts}. The second singular vector in \fg{fig:svd} is strikingly reminiscent of the carbon-source-specific singular vector that we \citep{Slavov_eth_grr} found to account for $11\%$ of the response to changes in growth rate; it peaks during LOC in the continuous glucose-limited cultures, but during HOC in the ethanol batch culture. This phase shift between gene-expression and oxygen-consumption suggests an intriguing hypothesis, namely that the shift in the phase of some genes might account for their carbon-source-specific growth rate responses \citep{Slavov_eth_grr}. The third singular vector (bottom panels in \fg{fig:svd}) is correlated to the oxygen consumption, similar to the first one, but shows more variance across the different conditions of metabolic synchrony. 


\subsubsection*{Metabolic Cycling can Account for the Carbon-source-specific Growth Rate Response Based on the EAP Mechanism}  
Previously we reported that the growth-rate (GR) responses of $1500$ genes, which do not depend on the nature of the carbon source and the nutrient limitation, correlate very strongly with their phase of peak expression in metabolically synchronized glucose-limited cultures  \citep{Slavov_eth_grr}.  Based on this correlation and on the  changes in the relative durations of the metabolic phases that we measured \citep{Slavov_eth_grr}, we suggested that the growth-rate (GR) response of asynchronous cultures can be explained simply by the GR changes in the fraction of cells in different phases of the metabolic and division cycles \citep{Slavov_eth_grr, Slavov_batch_ymc}. We refer to this dependence between the composition of an asynchronous culture and the population-average gene-expression of the culture as ensemble average over phases (EAP).
  We also previously observed that a smaller subset of genes have opposite growth rate responses in yeast grown on glucose or ethanol carbon source \citep{Slavov_eth_grr}. Using the gene expression across the metabolic cycle of a culture grown in ethanol medium and a culture grown in glucose medium, we sought to test the hypothesis suggested by the second singular vector in \fg{fig:svd}, namely whether the carbon-source-specific growth-rate responses can also be explained by the EAP mechanism, similar to the universal growth-rate response  \citep{Slavov_eth_grr}. This hypothesis would require that if a gene  with a carbon-source-specific growth-rate response peaks in HOC in ethanol, it peaks in LOC in glucose or vice versa.   
 
To identify genes expressed in different metabolic phases and test whether such genes have different GR slopes, we selected the genes whose corresponding gene-gene correlation vectors in glucose and in ethanol are negatively correlated  \citep{Slavov_2009}, whose expression profiles are periodic as determined by correlation analysis \citep{Slavov_batch_ymc}, and whose GR slopes have opposite signs in ethanol and in glucose carbon source  \citep{Slavov_eth_grr}. Clustering the resulting subset of genes indicated that indeed those genes shift phases (meaning that their peaks move from HOC to LOC or vice versa) between ethanol and glucose, \fg{fig:outPhase}. For each carbon source, the genes that peak during HOC have positive GR slopes in that carbon source. Conversely, the genes that peak in LOC have negative GR slopes in that carbon source.  Thus, the EAP mechanism can explain the carbon-source-dependent GR responses of these genes, confirming our hypothesis. 

The genes peaking in LOC in glucose and in HOC in ethanol are enriched for energy generation in the mitochondria. These genes include \emph{YAT1, YAT2, IDP2} that we  characterized as having carbon-source-dependent GR responses and being induced with growth-rate in ethanol but not in glucose  \citep{Slavov_eth_grr}. This set of genes does not include structural mitochondrial genes (those are not periodic in the batch culture, see \fg{fig:transcripts}). Thus, the phase shift in the expression of mitochondrial genes between ethanol and glucose carbon source pertains to using mitochondria for energy production from ethanol rather than making new mitochondria.     

\section*{Discussion}
Spontaneous respiratory oscillations in continuous cultures of \emph{Saccharomyces cerevisiae} have been known for decades \citep{ymc_1969, Kaspar_1969} and treated as different phenomena depending on the frequency of their synchronization. By gradually changing two parameters, biomass-density and growth-rate, we were able to: $(i)$ demonstrate gradual transition between the low and the high frequency respiratory cycles, suggesting common origin and biology, and $(ii)$ rigorously quantify the dependence among the growth parameters of the synchronized cultures, their frequency of respiratory cycling, cell division, and gene-expression patterns. Furthermore, we exploit the differences between respiratory oscillations under different conditions to collect data that enabled us to characterize genes expressed in different subpopulations either with the metabolic (cell growth) or with the cell division cycle. Such analysis is essential for distinguishing condition-invariant gene-gene correlations due to the timing of physiological processes occurring in single cells from correlations induced by condition-dependent synchronization offsets between subpopulations in synchronized cultures. We took advantage of yeast as a model organism \citep{Botstein2011} to rigorously characterize dynamics of cell growth and division that are conserved in other eukaryotes \citep{Slavov_emc} less amenable to such quantitative experiments.

\subsubsection*{Influence of Carbon-source, Growth-rate and Biomass-density on Gene Expression and Cell Division}
A major result of our data and analysis is that the vast majority of highly-periodic genes are correlated to each other in essentially the same way across all profiled conditions, regardless of strain-background, carbon-source,  growth-rate and biomass density; the mean-levels and the dynamics of gene-expression change substantially among the diverse conditions we studied, but the relationships between the dynamic patterns of gene expression remain condition invariant. This result strongly suggests that the fundamental biological functions of metabolism and cell growth are conserved across the studied strain backgrounds, (CEN.PK and \emph{S288c}), and actively regulated so that the gene correlations among growth genes are preserved even when the duration of the phases of metabolic cycling changes substantially (\fg{fig:CDC_YMC_periods}). 

We also found that depending on the culture conditions, cell division shifts relative to the metabolic phases. This result is supported both by the shift between the DNA-content and the oxygen consumption data (\fg{fig:dna_content}), and by the shifts between the gene expression levels of the genes oscillating with the two cycles (\fg{fig:cdc_div}A). The G1/S transition always occurs during or around the HOC phase (\fg{fig:dna_content}) but with a substantial variation depending on the growth conditions of metabolic synchrony. Our data do not address directly the factors determining this variation and future studies are required for its characterization. It could be that this variation is related to the interaction between the quorum-sensing mechanism, which entrains the period of cycling at the population level, and the average time that a cell needs, for the given experimental growth-conditions, to complete the metabolic phases and become ready for the G1/S transition. Similar temporal coincidence between high mitochondrial activity and oxygen consumption and the G1/S transition is also observed in \emph{Drosophila melanogaster} \citep{owusu2008distinct,Mandal22022010} and human cells \citep{schieke_coordination_2008, mitra_hyperfused_2009}, and may be a general regulatory mechanism \citep{Slavov_emc}.

\subsubsection*{The Frequency of Metabolic Cycling in Synchronized Cultures Depends on Their Biomass Density}
The constant level of residual glucose in \fg{fig:glu} reinforces our previous suggestion \citep{Slavov_batch_ymc} that oscillations in the concentration of glucose or another limiting nutrient are unlikely to directly drive  the oscillations in the synchronized cultures. We suggest that the glucose concentration in the feed media modulates the frequency of metabolic cycling of synchronized cultures (\fg{fig:CDC_YMC_periods}C) by modulating their biomass density; the higher the concentration of glucose in the feed media, the higher the biomass, and thus the higher the concentration of the chemical that mediates the synchrony, and the higher the frequency of cycling. This dependence is commonly observed in models of quorum-sensing based synchronization \citep{Strogatz_pnas_2004, schwab_dynamical_2010} and likely relevant to the metabolic synchronization as well. It may provide the unifying principle behind the ``short'' respiratory cycle of $45min$ \citep{klevecz_genomewide_2004} observed in IFO cultures fed with a medium containing high glucose concentration ($22 g/L$) and the ``longer'' cycles of $4-5 h$  \citep{tu_logic_2005} observed in CEN.PK cultures fed with a medium containing lower glucose concentration ($10 g/L$). The difference in frequency is also likely to be strain dependent, i.e., different sensitivity for, and/or a different rate of secretion of, the quorum sensing chemical mediating synchrony. In our experiments with a strain having S288c background, we observe a wide range of frequencies (\fg{fig:CDC_YMC_periods}C) at substantially lower concentrations of glucose in the feed media ($0.2 - 3.2 g/L$), corresponding to lower biomass densities.

In contrast to the diploid \emph{DBY12007} strain, its corresponding isogenic haploid strain did not synchronize in the same growth conditions at all growth rates.  The understanding of this negative result requires further experiments. We propose that this difference between the ability of haploid and diploid cells to synchronize metabolically is likely related to differences in the quorum sensing between haploid and diploid yeast.

\subsubsection*{Relationship between the CDC and Metabolic Cycling in Single Cells and in Synchronized Cultures}
We propose that the respiratory cycle observed in synchronized populations \citep{ymc_1969,Kaspar_1969} is emergent harmonic cycle of the metabolic/growth cycle in  single cells. In this view, our measurements in synchronized cultures can be explained as a superposition of two cycles, growth and division, at the single-cell level. What is the relationship between these cell autonomous cycles of growth and division in single-cells in the absence of cell-cell communication? In our culture with the lowest biomass density ($\mu=0.10h^{-1}$ and $[Glu]_{feed} = 300mg/L$), most cells go through a single LOC and a single HOC phase before dividing as indicated by the observation that $70\%$ of the cells divide during each period of the culture, \fg{fig:dna_content}. The extrapolation of this observation to the single-cell level predicts that a single-cell, when unperturbed, is likely to go through a single LOC and HOC phase, replicate its DNA and divide. In contrast, if the growth is entrained by cell-cell communication or perturbed by external stress, HOC and LOC can be shortened so that each cell has to go through multiple growth phases (as observed in the more dense cultures in \fg{fig:CDC_YMC_periods}C) before it is ready to enter the S phase. In this context, the cell growth/metabolic cycle and the CDC may be viewed not as separate cycles but as different phases of the cell life cycle that are ordered in a sequence by the cyclin-dependent-kinases and the checkpoints. Our data indicate that the relative duration and prominence of each phase changes, depending on the growth conditions, to optimize cell growth and division in a changing environment.

The G1/S transition occurs close to the HOC phase in all continuous cultures that we studied. This observation dovetails with similar observations across divergent eukaryotes \citep{owusu2008distinct,Mandal22022010, schieke_coordination_2008, mitra_hyperfused_2009,Slavov_emc} and brings about an attractive possibility: The activity of the target of rapamycin (TOR), a regulator of biosynthesis and respiration \citep{mTOR_respiration}, may coordinate the metabolic and division cycles. Indeed, the low activity of mammalian TOR observed in early G1 and high activity in late G1 and S/G2/M phases \citep{schieke_coordination_2008} can account for most changes in gene expression, translation rate and respiration level observed during metabolic cycling. The characterization of the molecular mechanisms by which TOR and other signalling pathways can control and be controlled by the cycles of growth and division remains to be fully characterized \citep{TOR-CDC-2004, duan2011linking}. 

High-precision measurements of cell-volume \citep{goranov2009rate}, cell-mass and cell-density \citep{bryan2010measurement} of single yeast cells  have characterized the  values and the variation of these cell parameters during the cell division cycle of yeast growing in rich media. These results may be related to the growth cycle that we study but there are two major differences worth emphasizing. First, the previous studies focused on cells growing in rich media, in the regime of aerobic glycolysis, and we have no direct data connecting the growth cycle that we observe in slowly growing, respiring yeast to the growth and physiology of yeast growing by aerobic glycolysis in rich media. The second major difference is in what is measured; the oscillations in storage carbohydrates are exactly out of phase with the expected oscillations in synthesizing cell biomass and the two may largely cancel their respective effects on the cell volume and mass, as would be suggested by the constant flux of glucose uptake, \fg{fig:glu}. Indeed, studies that have measured both protein synthesis and cell size (volume and mass) reported oscillations in protein synthesis but not in cell volume and size \citep{creanor1982patterns}. Based on the connection between the stress-response that we suggested \citep{Slavov_emc} and the correlation between mRNAs and their corresponding proteins observed during osmotic stress  \citep{Gasch2011dynamic}, we expect that many of the increases in mRNA levels will be reflected on the protein level as well albeit with altered magnitudes of change.

\subsubsection*{Connecting the Dynamics of Cell Growth to the Gene Expression of Exponentially-Growing Steady-State Asynchronous Cultures}
We previously quantified the growth-rate dependent changes in the relative durations of the HOC and the LOC phases based on the dissolved oxygen data in the media of synchronized cultures \citep{Slavov_eth_grr}.  Since such growth-rate changes in the single-cell autonomous cell growth/metabolic cycle will affect the composition of an asynchronous culture (fraction of cells in different phases of the growth cycle), we have suggested that the population-average gene-expression of the culture is an ensemble average over the phases (EAP) of the constituent cells \citep{Slavov_eth_grr, Slavov_batch_ymc, Slavov_emc}. Our new data reinforce and extend this model in two directions. First, our gene expression data demonstrate that the changes in the relative durations of the LOC and the HOC phase occur not only at the level of oxygen consumption but extend to the vast majority of periodically expressed genes (\fg{fig:transcripts}) as the model both predicts and requires. Second, we demonstrate that the EAP mechanism can account for the population-average expression-levels not only for the genes with  carbon-source-independent growth-rate responses but also for genes with  carbon-source-dependent growth-rate responses. This finding provides further support for the EAP mechanism and suggests that changes in the carbon-source alter the timing of expression of hundreds of genes both in synchronous (\fg{fig:outPhase}) and in asynchronous cultures \citep{Slavov_eth_grr}. Taken together, our data and analysis underscore that cell growth is not constant throughout the cell life-cycle but dynamically coupled to cell division in response to nutrient availability and other conditions. Thus, understanding cell growth and division at the single-cell level requires that both experiments and analysis transcend the confines of steady-state approximations and delve into the rich dynamics of the physiological processes that define the life cycle of cells.

\section*{Materials and Methods}
\subsection*{Growth Conditions} 
We synchronized metabolically a diploid prototrophic strain (\emph{DBY12007}) with \emph{S288c} background and WT \emph{HAP1}, originally derived by \citet{hickman2007heme}, using starvation followed by re-feeding. The media was limited by glucose with composition described by \citet{Slavov_eth_grr}.  Chemostats were established in $500mL$ fermenter vessels (Sixfors; Infors AG, Bottmingen, Switzerland) containing $300 mL$ of culture volume, stirred at $400 rpm$, and aerated with humidified and filtered air.

\subsection*{Measuring mRNA} 
To measure mRNA levels, we sampled \emph{2-4ml} of stably cycling culture, from the effluent and without disturbing the culture, and vacuum filtered the cells, followed by immediate freezing in liquid nitrogen and then in a freezer at $-80^oC$. RNA for microarray analysis was extracted by the acid--phenol--chloroform method. RNA was amplified and labeled using the Agilent low RNA input fluorescent linear amplification kit (P/N 5184-3523; Agilent Technologies, Palo Alto, CA). This method involves initial synthesis of cDNA by using a poly(T) primer attached to a T7 promoter. Labeled cRNA is subsequently synthesized using T7 RNA polymerase and either \emph{Cy3} or \emph{Cy5} UTP. Each \emph{Cy5}-labeled experimental cRNA sample was mixed with the \emph{Cy3}-labeled reference cRNA and hybridized for 17 h at $65^oC$ to custom Agilent Yeast oligo microarrays $8 \times 15k$ having $8$ microarrays per glass slide. Microarrays were washed, scanned with an Agilent DNA microarray scanner (Agilent Technologies), and the resulting TIF files processed using Agilent Feature Extraction Software version $7.5$. Resulting microarray intensity data were submitted to the \href{http://puma.princeton.edu}{PUMA Database} for archiving. 

Results for each gene and time point were expressed as the $\log_2$ of the sample signal divided by the signal in the reference channel. Genes whose expression vectors contained more than $20\%$ missing values were filtered out. The small number of missing values in the expression vectors of the remaining genes were replaced by the values computed from the linear interpolation of the neighboring datapoints. 

\subsection*{Measuring DNA Content} 
To measure DNA content, we diluted $200\mu l$ culture from each sampled time point with ethanol to a final concentration of $70\%$ ethanol and stored the samples in a freezer at $-20^oC$. Subsequently, the cells from each sample were pelleted and washed in $800 \mu l$ $50 mM$ sodium citrate at pH $7.2$. The washed cells were incubated overnight at $50^oC$ in $500 \mu l$ sodium citrate containing $250 \mu g/ml$ RNase A. Next, $50 \mu l$ of $20 mg/ml $ proteinase K were added to each sample and the samples were returned back to $50^oC$ for $2$ more hours. After the incubation, each sample was sonicated for $10s$ and stained by the addition of $500 \mu l$ $50 mM$ sodium citrate containing $3 \mu M$ SYTOX Green (stock is 5 mM in DMSO from Molecular Probes Invitrogen). After $1$ hour incubation in dark, the DNA content of single cells from each sample was quantified by fluorescence activated cell sorting (FACS).

\subsection*{Measuring Residual Glucose}
Residual glucose was assayed in the filtrate (growth medium) from the samples for mRNA by enzyme-coupled NADH oxidation reactions (assay kits from
R-Biopharm, Darmstadt, Germany). Between the filtration and the measurement, the samples were stored at $-20^oC$.

\subsection*{Quantifying the period of metabolic cycling}
To quantify the period of a metabolically synchronized culture, we used both ($i$) Fourier analysis and ($ii$) direct measurements of the time-interval between two successive time-points during which the culture has the same metabolic phase. For higher accuracy, we used the oxygen consumption data only during stable oscillations (the limit cycle regimes) and averaged data from multiple cycles.  The two approaches gave statistically indistinguishable results and we present the result from the more direct and intuitive measurement ($ii$).

\section*{Acknowledgments}
We thank Sanford J. Silverman, David Schwab, and Jennifer Ewald for stimulating discussions and feedback on the manuscript. Research was funded by a grant from the National Institutes of Health (GM046406) and the NIGMS Center for Quantitative Biology (GM071508). 

\section*{Figure Captions}

\begin{figure}[!ht]
 \caption{ 
  {\bf The period of the metabolic cycling in synchronized cultures depends on the growth rate and on the biomass density.} 
  (A) The metabolic cycling can be can be divided phenomenologically into two phases, the low oxygen consumption phase (LOC), when the amount of oxygen in the media is high because the cells consume little oxygen, and the high oxygen consumption phase (HOC) when the reverse holds.  
  (B) The HOC and LOC phases can be quantified objectively based on the bimodality of the distribution of dissolved oxygen.
  (C) Independent continuous glucose-limited cultures were metabolically synchronized at the doubling periods indicated on the x-axis and different biomass-densities corresponding to the concentrations of glucose in the feed media indicated in the legend. 
 }		
 \label{fig:CDC_YMC_periods}	
\end{figure}

\begin{figure}[!ht]
 \caption{ 
  {\bf The residual glucose concentration remains constant throughout the metabolic cycling of synchronized cultures.}
  Concentration of glucose in the culture media of a metabolically synchronized culture at growth rate $\mu = 0.133 h^{-1}$ and glucose concentration in the feed media $[Glucose] = 800mg/L$. The dissolved oxygen in the media is shown on the top.  
 }			
 \label{fig:glu}	
\end{figure}

\begin{figure}[h!!!!]	
 \caption{ 
 {\bf The CDC phases shift relative to the phases of the oxygen consumption.}
 Four cultures were synchronized at different growth rates and carbon-sources (top) and profiled for   
   DNA content and oxygen consumption.  (A) The distributions of DNA content in single cell were measured by FACS and used to infer the fraction of cells in the phases of the CDC (B). 
 }			
 \label{fig:dna_content}		
\end{figure}	

\begin{figure}[!ht]
 \caption{ 
  {\bf Clustered gene-expression data from five cultures synchronized at different growth conditions.}
  Hierarchically clustered gene expression data from the cultures described in \fg{fig:dna_content} and from a continuous glucose-limited culture \citep{tu_logic_2005}. The data from each culture and for each gene has been centered to zero mean to emphasize the oscillatory pattern. The dissolved oxygen in the media during the sampling, scaled for best visibility, is indicated by the bars on the top. Clustering is based on non-centered correlation computed using all shown data. The gene-expression data in this and all other figures are displayed on a $\log_2$ scale \citep{eisen1998cluster}. 
 } 
\label{fig:transcripts}		
\end{figure}

\begin{figure}[!ht]
 \caption{ 
  {\bf The expression levels of genes annotated to cell division correlate either to oxygen consumption or to DNA content.} 
  The genes annotated to the CDC can be separated into two subsets based on their expression levels: genes correlating to DNA content and this cell division (A) and genes correlating to oxygen consumption (B).   
  The top panels show the fraction of cells with duplicated DNA in each culture from \fg{fig:dna_content}, rescaled for better visibility. The middle panels show the dissolved oxygen, also rescaled for visibility. The bottom panels show the gene expression levels, displayed on a $\log_2$ scale, for the subsets of genes annotated to the CDC that either correlate to cell division (A) or to oxygen consumption (B).     
 }
 \label{fig:cdc_div}	
\end{figure}

\begin{figure}[!ht]
 \caption{ 
  {\bf Dominant gene expression patterns.}
  Singular value decomposition (SVD) of the gene expression data from \fg{fig:transcripts}. Each column corresponds to a synchronized culture, the same order an notation as in \fg{fig:transcripts} and each row corresponds to a singular vector. The fraction of the variance $\sigma^2$ explained by each singular vector is indicated by the red numbers to the right of the panels.     	 
 } 
\label{fig:svd}		
\end{figure}

\begin{figure}[!ht]
 \caption{ {\bf Hundreds of periodic genes shift phases between glucose and ethanol carbon-sources.}
The mean-centered expression levels of genes with phase-shifts are shown in the phosphate-limited ethanol batch culture (left panel) and in the glucose-limited continuous culture (middle panel) growing at $\mu=0.095 h^{-1}$. The right panel shows the growth rate (GR) slopes from the same genes in phosphate-limited ethanol media (first column) and in glucose-limited media (second column). The clustering permutation was computed based only on the gene expression data from the metabolically synchronized cultures. We previously computed growth rate (GR) slopes by regressing transcript levels against the growth rate \citep{brauer_2008,Slavov_eth_grr}; positive GR slopes indicate increasing expression with growth rate and negative GR slopes indicate decreasing expression. }	
\label{fig:outPhase}
\end{figure}
\cleardoublepage

\bibliographystyle{C:/Users/nslavov/Documents/B/latex/msb} 
\bibliography{C:/Users/nslavov/Papers/REFs/pops,C:/Users/nslavov/Papers/REFs/dissertation,C:/Users/nslavov/Papers/REFs/grr,C:/Users/nslavov/Papers/REFs/EthCyc,C:/Users/nslavov/Papers/REFs/auxo_1,C:/Users/nslavov/Papers/REFs/aerobic_glycolysis,C:/Users/nslavov/Papers/REFs/emc,C:/Users/nslavov/Papers/REFs/people/my,C:/Users/nslavov/Papers/REFs/people/JM_Michison,C:/Users/nslavov/Papers/REFs/people/people}

\begin{thebibliography}{71}
\expandafter\ifx\csname natexlab\endcsname\relax\def\natexlab#1{#1}\fi

\bibitem[{Benanti(2012)}]{benanti2012coordination}
Benanti J (2012) Coordination of Cell Growth and Division by the
  Ubiquitin-Proteasome System. In {\it Seminars in Cell \& Developmental
  Biology\/}. Elsevier

\bibitem[{Blank {\it et~al\/}(2009)Blank, Gajjar, Belyanin \&
  Polymenis}]{Polymenis-mtDNA-sulfur}
Blank HM, Gajjar S, Belyanin A, Polymenis M (2009) Sulfur Metabolism Actively
  Promotes Initiation of Cell Division in Yeast. {\it PLoS ONE\/} {\bf 4}:
  e8018

\bibitem[{Boer {\it et~al\/}(2008)Boer, Amini \&
  Botstein}]{boer_influence_2008}
Boer VM, Amini S, Botstein D (2008) Influence of genotype and nutrition on
  survival and metabolism of starving yeast. {\it Proceedings of the National
  Academy of Sciences\/} {\bf 105}: 6930--6935

\bibitem[{Botstein and Fink(2011)Botstein \& Fink}]{Botstein2011}
Botstein D, Fink GR (2011) {Yeast: an experimental organism for 21st century
  biology.} {\it Genetics\/} {\bf 189}: 695--704

\bibitem[{Brauer {\it et~al\/}(2008)Brauer, Huttenhower, Airoldi, Rosenstein,
  Matese, Gresham, Boer, Troyanskaya \& Botstein}]{brauer_2008}
Brauer MJ, Huttenhower C, Airoldi EM, Rosenstein R, Matese JC, Gresham D, Boer
  VM, Troyanskaya OG, Botstein D (2008) Coordination of Growth Rate, Cell
  Cycle, Stress Response, and Metabolic Activity in Yeast. {\it Mol Biol
  Cell\/} {\bf 19}: 352--367

\bibitem[{Bryan {\it et~al\/}(2010)Bryan, Goranov, Amon \&
  Manalis}]{bryan2010measurement}
Bryan A, Goranov A, Amon A, Manalis S (2010) Measurement of mass, density, and
  volume during the cell cycle of yeast. {\it Proceedings of the National
  Academy of Sciences\/} {\bf 107}: 999

\bibitem[{Cairns {\it et~al\/}(2011)Cairns, Harris \&
  Mak}]{cairns_regulation_2011}
Cairns RA, Harris IS, Mak TW (2011) Regulation of cancer cell metabolism. {\it
  Nature Reviews Cancer\/} {\bf 11}: 85--95, {PMID:} 21258394

\bibitem[{Chen and Fink(2006)Chen \& Fink}]{2006_quorum_fink}
Chen H, Fink G (2006) Feedback control of morphogenesis in fungi by aromatic
  alcohols. {\it Genes  development\/} {\bf 20}: 1150

\bibitem[{Chen {\it et~al\/}(2004)Chen, Fujita, Feng, Clardy \&
  Fink}]{2004_quorum_fink_Candida_albicans}
Chen H, Fujita M, Feng Q, Clardy J, Fink G (2004) Tyrosol is a quorum-sensing
  molecule in Candida albicans. {\it Proceedings of the National Academy of
  Sciences of the United States of America\/} {\bf 101}: 5048

\bibitem[{Chin {\it et~al\/}(2012)Chin, Marcus, Klevecz \&
  Li}]{genome-wide-transcriptional-oscillators}
Chin SL, Marcus IM, Klevecz RR, Li CM (2012) Dynamics of oscillatory phenotypes
  in \emph{Saccharomyces cerevisiae} reveal a network of genome-wide
  transcriptional oscillators. {\it FEBS Journal\/}

\bibitem[{Creanor and Mitchison(1982)Creanor \&
  Mitchison}]{creanor1982patterns}
Creanor J, Mitchison J (1982) Patterns of protein synthesis during the cell
  cycle of the fission yeast Schizosaccharomyces pombe. {\it Journal of cell
  science\/} {\bf 58}: 263--285

\bibitem[{Dang(2012)}]{dang2012metabolism_and_cancer}
Dang C (2012) Links between metabolism and cancer. {\it Genes  Development\/}
  {\bf 26}: 877--890

\bibitem[{Duan and Pagano(2011)Duan \& Pagano}]{duan2011linking}
Duan S, Pagano M (2011) Linking metabolism and cell cycle progression via the
  APC/CCdh1 and SCF$\beta$TrCP ubiquitin ligases. {\it Proceedings of the
  National Academy of Sciences\/} {\bf 108}: 20857--20858

\bibitem[{Eisen {\it et~al\/}(1998)Eisen, Spellman, Brown \&
  Botstein}]{eisen1998cluster}
Eisen M, Spellman P, Brown P, Botstein D (1998) Cluster analysis and display of
  genome-wide expression patterns. {\it Proceedings of the National Academy of
  Sciences\/} {\bf 95}: 14863

\bibitem[{Ferea {\it et~al\/}(1999)Ferea, Botstein, Brown \&
  Rosenzweig}]{ferea1999systematic}
Ferea T, Botstein D, Brown P, Rosenzweig R (1999) Systematic changes in gene
  expression patterns following adaptive evolution in yeast. {\it Proceedings
  of the National Academy of Sciences\/} {\bf 96}: 9721

\bibitem[{Fingar and Blenis(2004{\natexlab{a}})Fingar \&
  Blenis}]{fingar2004target}
Fingar D, Blenis J (2004{\natexlab{a}}) Target of rapamycin (TOR): an
  integrator of nutrient and growth factor signals and coordinator of cell
  growth and cell cycle progression. {\it Oncogene\/} {\bf 23}: 3151--3171

\bibitem[{Fingar and Blenis(2004{\natexlab{b}})Fingar \& Blenis}]{TOR-CDC-2004}
Fingar D, Blenis J (2004{\natexlab{b}}) Target of rapamycin (TOR): an
  integrator of nutrient and growth factor signals and coordinator of cell
  growth and cell cycle progression. {\it Oncogene\/} {\bf 23}: 3151--3171

\bibitem[{{Garcia-Ojalvo} {\it et~al\/}(2004){Garcia-Ojalvo}, Elowitz \&
  Strogatz}]{Strogatz_pnas_2004}
{Garcia-Ojalvo} J, Elowitz MB, Strogatz SH (2004) Modeling a synthetic
  multicellular clock: Repressilators coupled by quorum sensing. {\it
  Proceedings of the National Academy of Sciences of the United States of
  America\/} {\bf 101}: 10955 --10960

\bibitem[{Goranov and Amon(2010)Goranov \& Amon}]{goranov2010growth}
Goranov A, Amon A (2010) Growth and division--not a one-way road. {\it Current
  opinion in cell biology\/} {\bf 22}: 795--800

\bibitem[{Goranov {\it et~al\/}(2009)Goranov, Cook, Ricicova, Ben-Ari,
  Gonzalez, Hansen, Tyers \& Amon}]{goranov2009rate}
Goranov A, Cook M, Ricicova M, Ben-Ari G, Gonzalez C, Hansen C, Tyers M, Amon A
  (2009) The rate of cell growth is governed by cell cycle stage. {\it Genes
  development\/} {\bf 23}: 1408

\bibitem[{Hickman and Winston(2007)Hickman \& Winston}]{hickman2007heme}
Hickman M, Winston F (2007) Heme levels switch the function of {Hap1} of {{\it
  Saccharomyces cerevisiae}} between transcriptional activator and
  transcriptional repressor. {\it Molecular and Cellular Biology\/} {\bf 27}:
  7414--7424

\bibitem[{Hoose {\it et~al\/}(2012{\natexlab{a}})Hoose, Rawlings, Kelly,
  Leitch, Ababneh, Robles, Taylor, Hoover, Hailu, McEnery {\it
  et~al\/}}]{hoose2012systematic}
Hoose S, Rawlings J, Kelly M, Leitch M, Ababneh Q, Robles J, Taylor D, Hoover
  E, Hailu B, McEnery K, {\it et~al\/} (2012{\natexlab{a}}) A Systematic
  Analysis of Cell Cycle Regulators in Yeast Reveals That Most Factors Act
  Independently of Cell Size to Control Initiation of Division. {\it PLoS
  Genetics\/} {\bf 8}: e1002590

\bibitem[{Hoose {\it et~al\/}(2012{\natexlab{b}})Hoose, Rawlings, Kelly,
  Leitch, Ababneh, Robles, Taylor, Hoover, Hailu, McEnery, Downing, Kaushal,
  Chen, Rife, Brahmbhatt, Smith \& Polymenis}]{Polymenis-division-control}
Hoose SA, Rawlings JA, Kelly MM, Leitch MC, Ababneh QO, Robles JP, Taylor D,
  Hoover EM, Hailu B, McEnery KA, Downing SS, Kaushal D, Chen Y, Rife A,
  Brahmbhatt KA, Smith III R, Polymenis M (2012{\natexlab{b}}) A Systematic
  Analysis of Cell Cycle Regulators in Yeast Reveals That Most Factors Act
  Independently of Cell Size to Control Initiation of Division. {\it PLoS
  Genet\/} {\bf 8}: e1002590

\bibitem[{Hornby {\it et~al\/}(2001)Hornby, Jensen, Lisec, Tasto, Jahnke,
  Shoemaker, Dussault \& Nickerson}]{hornby2001quorum}
Hornby J, Jensen E, Lisec A, Tasto J, Jahnke B, Shoemaker R, Dussault P,
  Nickerson K (2001) Quorum sensing in the dimorphic fungus Candida albicans is
  mediated by farnesol. {\it Applied and environmental microbiology\/} {\bf
  67}: 2982

\bibitem[{Jain {\it et~al\/}(2012)Jain, Nilsson, Sharma, Madhusudhan, Kitami,
  Souza, Kafri, Kirschner, Clish \& Mootha}]{Jain25052012}
Jain M, Nilsson R, Sharma S, Madhusudhan N, Kitami T, Souza AL, Kafri R,
  Kirschner MW, Clish CB, Mootha VK (2012) Metabolite Profiling Identifies a
  Key Role for Glycine in Rapid Cancer Cell Proliferation. {\it Science\/} {\bf
  336}: 1040--1044

\bibitem[{Johnston {\it et~al\/}(1977)Johnston, Pringle \&
  Hartwell}]{johnston1977coordination}
Johnston G, Pringle J, Hartwell L (1977) Coordination of growth with cell
  division in the yeast Saccharomyces cerevisiae. {\it Experimental cell
  research\/} {\bf 105}: 79--98

\bibitem[{Jorgensen {\it et~al\/}(2002)Jorgensen, Nishikawa, Breitkreutz \&
  Tyers}]{jorgensen2002systematic}
Jorgensen P, Nishikawa J, Breitkreutz B, Tyers M (2002) Systematic
  identification of pathways that couple cell growth and division in yeast.
  {\it Sciences STKE\/} {\bf 297}: 395

\bibitem[{Jorgensen and Tyers(2004)Jorgensen \& Tyers}]{jorgensen2004cells}
Jorgensen P, Tyers M (2004) How cells coordinate growth and division. {\it
  Current Biology\/} {\bf 14}: R1014--R1027

\bibitem[{Keulers {\it et~al\/}(1996)Keulers, Satroutdinov, Suzuki \&
  Kuriyama}]{keulers1996synchronization_CO2}
Keulers M, Satroutdinov A, Suzuki T, Kuriyama H (1996) Synchronization affector
  of autonomous short-period-sustained oscillation of Saccharomyces cerevisiae.
  {\it Yeast\/} {\bf 12}: 673--682

\bibitem[{Klevecz {\it et~al\/}(2004)Klevecz, Bolen, Forrest \&
  Murray}]{klevecz_genomewide_2004}
Klevecz RR, Bolen J, Forrest G, Murray DB (2004) A genomewide oscillation in
  transcription gates {DNA} replication and cell cycle. {\it Proceedings of the
  National Academy of Sciences of the United States of America\/} {\bf 101}:
  1200--1205

\bibitem[{K\"{u}enzi and Fiechter(1969)K\"{u}enzi \& Fiechter}]{ymc_1969}
K\"{u}enzi MT, Fiechter A (1969) Changes in carbohydrate composition and
  trehalase-activity during the budding cycle of Saccharomyces cerevisiae. {\it
  Archiv Fr Mikrobiologie\/} {\bf 64}: 396--407, {PMID:} 4916776

\bibitem[{Kwak {\it et~al\/}(2003)Kwak, Kwon, Jin, Kuriyama \&
  Sohn}]{Kuriyama2003_H2S}
Kwak W, Kwon G, Jin I, Kuriyama H, Sohn H (2003) Involvement of oxidative
  stress in the regulation of H2S production during ultradian metabolic
  oscillation of Saccharomyces cerevisiae. {\it FEMS microbiology letters\/}
  {\bf 219}: 99--104

\bibitem[{Laxman and Tu(2010)Laxman \& Tu}]{Laxman2010}
Laxman S, Tu BP (2010) {Systems approaches for the study of metabolic cycles in
  yeast}. {\it Current Opinion in Genetics  Development\/} {\bf 20}: 599--604

\bibitem[{Lee {\it et~al\/}(2011)Lee, Topper, Hubler, Hose, Wenger, Coon \&
  Gasch}]{Gasch2011dynamic}
Lee M, Topper S, Hubler S, Hose J, Wenger C, Coon J, Gasch A (2011) A dynamic
  model of proteome changes reveals new roles for transcript alteration in
  yeast. {\it Molecular systems biology\/} {\bf 7}

\bibitem[{Machn{\'e} and Murray(2012)Machn{\'e} \& Murray}]{Murray2012yin}
Machn{\'e} R, Murray D (2012) The Yin and Yang of Yeast Transcription: Elements
  of a Global Feedback System between Metabolism and Chromatin. {\it PloS
  one\/} {\bf 7}: e37906

\bibitem[{Mandal {\it et~al\/}(2010)Mandal, Freije, Guptan \&
  Banerjee}]{Mandal22022010}
Mandal S, Freije WA, Guptan P, Banerjee U (2010) Metabolic control of G1-S
  transition: cyclin E degradation by p53-induced activation of the
  ubiquitin-proteasome system. {\it The Journal of Cell Biology\/} {\bf 188}:
  473--479

\bibitem[{Mitchison(1969)}]{mitchison1969enzyme}
Mitchison J (1969) Enzyme synthesis in synchronous cultures. {\it Science\/}
  {\bf 165}: 657

\bibitem[{Mitra {\it et~al\/}(2009)Mitra, Wunder, Roysam, Lin \&
  {Lippincott-Schwartz}}]{mitra_hyperfused_2009}
Mitra K, Wunder C, Roysam B, Lin G, {Lippincott-Schwartz} J (2009) A hyperfused
  mitochondrial state achieved at {G1-S} regulates cyclin E buildup and entry
  into S phase. {\it Proceedings of the National Academy of Sciences of the
  United States of America\/} {\bf 106}: 11960--11965, {PMID:} 19617534

\bibitem[{Murray {\it et~al\/}(2007)Murray, Beckmann \&
  Kitano}]{murray2007regulation}
Murray D, Beckmann M, Kitano H (2007) Regulation of yeast oscillatory dynamics.
  {\it Proceedings of the National Academy of Sciences\/} {\bf 104}: 2241

\bibitem[{Murray {\it et~al\/}(1999)Murray, Engelen, Lloyd \&
  Kuriyama}]{murray1999glutathione}
Murray D, Engelen F, Lloyd D, Kuriyama H (1999) Involvement of glutathione in
  the regulation of respiratory oscillation during a continuous culture of
  Saccharomyces cerevisiae. {\it Microbiology\/} {\bf 145}: 2739--2745

\bibitem[{Murray {\it et~al\/}(2003)Murray, Klevecz \&
  Lloyd}]{murray2003acetaldehyde_sulfite}
Murray D, Klevecz R, Lloyd D (2003) Generation and maintenance of synchrony in
  Saccharomyces cerevisiae continuous culture. {\it Experimental cell
  research\/} {\bf 287}: 10--15

\bibitem[{Murray {\it et~al\/}(2011)Murray, Haynes \&
  Tomita}]{murray_redox_2011}
Murray DB, Haynes K, Tomita M (2011) Redox regulation in respiring
  Saccharomyces cerevisiae. {\it Biochimica Et Biophysica Acta\/} {\bf 1810}:
  945--958, {PMID:} 21549177

\bibitem[{Murray and Lloyd(2007)Murray \& Lloyd}]{murray_tuneable_2007}
Murray DB, Lloyd D (2007) A tuneable attractor underlies yeast respiratory
  dynamics. {\it Bio Systems\/} {\bf 90}: 287--294, {PMID:} 17074432

\bibitem[{Neufeld and Edgar(1998)Neufeld \& Edgar}]{neufeld1998connections}
Neufeld T, Edgar B (1998) Connections between growth and the cell cycle. {\it
  Current opinion in cell biology\/} {\bf 10}: 784--790

\bibitem[{Owusu-Ansah {\it et~al\/}(2008)Owusu-Ansah, Yavari, Mandal \&
  Banerjee}]{owusu2008distinct}
Owusu-Ansah E, Yavari A, Mandal S, Banerjee U (2008) Distinct mitochondrial
  retrograde signals control the G1-S cell cycle checkpoint. {\it Nature
  genetics\/} {\bf 40}: 356--361

\bibitem[{Polymenis and Schmidt(1997)Polymenis \&
  Schmidt}]{polymenis1997coupling}
Polymenis M, Schmidt E (1997) Coupling of cell division to cell growth by
  translational control of the G1 cyclin CLN3 in yeast. {\it Genes
  development\/} {\bf 11}: 2522

\bibitem[{Robertson {\it et~al\/}(2008)Robertson, Stowers, Boczko \&
  Johnson}]{robertson_real-time_2008}
Robertson JB, Stowers CC, Boczko E, Johnson CH (2008) Real-time luminescence
  monitoring of cell-cycle and respiratory oscillations in yeast. {\it
  Proceedings of the National Academy of Sciences\/} {\bf 105}: 17988 --17993

\bibitem[{Ronen and Botstein(2006)Ronen \&
  Botstein}]{ronen_transcriptional_2006}
Ronen M, Botstein D (2006) Transcriptional response of steady-state yeast
  cultures to transient perturbations in carbon source. {\it Proceedings of the
  National Academy of Sciences of the United States of America\/} {\bf 103}:
  389--394

\bibitem[{Rousset {\it et~al\/}(1979)Rousset, Chevalier, Rousset, Dussaulx \&
  Zweibaum}]{rousset1979presence}
Rousset M, Chevalier G, Rousset J, Dussaulx E, Zweibaum A (1979) Presence and
  cell growth-related variations of glycogen in human colorectal adenocarcinoma
  cell lines in culture. {\it Cancer research\/} {\bf 39}: 531

\bibitem[{Sasidharan {\it et~al\/}(2012)Sasidharan, Tomita, Aon, Lloyd \&
  Murray}]{2012_Murray}
Sasidharan K, Tomita M, Aon M, Lloyd D, Murray DB (2012) Time-Structure of the
  Yeast Metabolism In vivo, In {\it Advances in Systems Biology\/}, Goryanin
  II, Goryachev AB (eds), volume 736 of {\it Advances in Experimental Medicine
  and Biology\/},  Springer New York, ISBN 978-1-4419-7210-1, pp. 359--379

\bibitem[{Schieke {\it et~al\/}(2008)Schieke, {McCoy} \&
  Finkel}]{schieke_coordination_2008}
Schieke SM, {McCoy} JP, Finkel T (2008) Coordination of mitochondrial
  bioenergetics with G1 phase cell cycle progression. {\it Cell Cycle
  Georgetown Tex\/} {\bf 7}: 1782--1787, {PMID:} 18583942

\bibitem[{Schieke {\it et~al\/}(2006)Schieke, Phillips, McCoy, Aponte, Shen,
  Balaban \& Finkel}]{mTOR_respiration}
Schieke SM, Phillips D, McCoy JP, Aponte AM, Shen RF, Balaban RS, Finkel T
  (2006) The Mammalian Target of Rapamycin (mTOR) Pathway Regulates
  Mitochondrial Oxygen Consumption and Oxidative Capacity. {\it Journal of
  Biological Chemistry\/} {\bf 281}: 27643--27652

\bibitem[{Schwab {\it et~al\/}(2010)Schwab, Baetica \&
  Mehta}]{schwab_dynamical_2010}
Schwab DJ, Baetica A, Mehta P (2010) Dynamical quorum-sensing and
  synchronization of nonlinear oscillators coupled through an external medium.
  {\it arXiv10124863\/}

\bibitem[{Shi {\it et~al\/}(2010)Shi, Sutter, Ye \& Tu}]{trehalose_2010}
Shi L, Sutter BM, Ye X, Tu BP (2010) Trehalose Is a Key Determinant of the
  Quiescent Metabolic State That Fuels Cell Cycle Progression upon Return to
  Growth. {\it Mol Biol Cell\/} {\bf 21}: 1982--1990

\bibitem[{Silverman {\it et~al\/}(2010)Silverman, Petti, Slavov, Parsons,
  Briehof, Thiberge, Zenklusen, Gandhi, Larson, Singer {\it
  et~al\/}}]{ymc_2010}
Silverman S, Petti A, Slavov N, Parsons L, Briehof R, Thiberge S, Zenklusen D,
  Gandhi S, Larson D, Singer R, {\it et~al\/} (2010) Metabolic cycling in
  single yeast cells from unsynchronized steady-state populations limited on
  glucose or phosphate. {\it Proceedings of the National Academy of Sciences\/}
  {\bf 107}: 6946--6951

\bibitem[{Slavov(2010)}]{slavov_thesis}
Slavov N (2010) {\it Universality, specificity and regulation of {\it S.
  cerevisiae} growth rate response in different carbon sources and nutrient
  limitations\/}. Ph.D. thesis, Princeton University

\bibitem[{Slavov {\it et~al\/}(2012)Slavov, Airoldi, van Oudenaarden \&
  Botstein}]{Slavov_emc}
Slavov N, Airoldi EM, van Oudenaarden A, Botstein D (2012) A conserved cell
  growth cycle can account for the environmental stress responses of divergent
  eukaryotes. {\it Molecular Biology of the Cell\/} {\bf 23}: 1986--1997

\bibitem[{Slavov and Botstein(2011)Slavov \& Botstein}]{Slavov_eth_grr}
Slavov N, Botstein D (2011) Coupling among growth rate response, metabolic
  cycle, and cell division cycle in yeast. {\it Molecular Biology of the
  Cell\/} {\bf 22}: 1997--2009

\bibitem[{Slavov and Botstein(2013)Slavov \& Botstein}]{Slavov_aux}
Slavov N, Botstein D (2013) {Decoupling Nutrient Signaling from Growth Rate
  Causes Aerobic Glycolysis and Deregulation of Cell-Size and Gene Expression}.
  {\it Molecular Biology of the Cell\/} {\bf 24}: 157--168

\bibitem[{Slavov {\it et~al\/}(2014)Slavov, Budnik, Schwab, Airoldi \& van
  Oudenaarden}]{Slavov_exp}
Slavov N, Budnik B, Schwab D, Airoldi E, van Oudenaarden A (2014) {Constant
  Growth Rate Can Be Supported by Decreasing Energy Flux and Increasing Aerobic
  Glycolysis}. {\it Cell Reports\/} {\bf 7}: 705 -- 714

\bibitem[{Slavov and Dawson(2009)Slavov \& Dawson}]{Slavov_2009}
Slavov N, Dawson KA (2009) {Correlation signature of the macroscopic states of
  the gene regulatory network in cancer}. {\it Proceedings of the National
  Academy of Sciences\/} {\bf 106}: 4079--4084

\bibitem[{Slavov {\it et~al\/}(2011)Slavov, Macinskas, Caudy \&
  Botstein}]{Slavov_batch_ymc}
Slavov N, Macinskas J, Caudy A, Botstein D (2011) Metabolic cycling without
  cell division cycling in respiring yeast. {\it Proceedings of the National
  Academy of Sciences of the United States of America\/} {\bf 108}:
  19090--19095

\bibitem[{Spellman {\it et~al\/}(1998)Spellman, Sherlock, Zhang, Iyer, Anders,
  Eisen, Brown, Botstein \& Futcher}]{spellman_1998}
Spellman PT, Sherlock G, Zhang MQ, Iyer VR, Anders K, Eisen MB, Brown PO,
  Botstein D, Futcher B (1998) Comprehensive Identification of Cell
  Cycle-regulated Genes of the Yeast Saccharomyces cerevisiae by Microarray
  Hybridization. {\it Mol Biol Cell\/} {\bf 9}: 3273--3297

\bibitem[{Tu and McKnight(2006)Tu \& McKnight}]{tu2006_review}
Tu B, McKnight S (2006) Metabolic cycles as an underlying basis of biological
  oscillations. {\it Nature Reviews Molecular Cell Biology\/} {\bf 7}: 696--701

\bibitem[{Tu and McKnight(2009)Tu \& McKnight}]{tu2009evidence_CO2}
Tu B, McKnight S (2009) Evidence of carbon monoxide-mediated phase advancement
  of the yeast metabolic cycle. {\it Proceedings of the National Academy of
  Sciences\/} {\bf 106}: 14293--14296

\bibitem[{Tu {\it et~al\/}(2005)Tu, Kudlicki, Rowicka \&
  {McKnight}}]{tu_logic_2005}
Tu BP, Kudlicki A, Rowicka M, {McKnight} SL (2005) Logic of the Yeast Metabolic
  Cycle: Temporal Compartmentalization of Cellular Processes. {\it Science\/}
  {\bf 310}: 1152--1158

\bibitem[{Turner {\it et~al\/}(2012)Turner, Ewald \& Skotheim}]{Jan_cell_size}
Turner J, Ewald J, Skotheim J (2012) Cell Size Control in Yeast. {\it Current
  biology\/} {\bf 22}: R350 -- R359

\bibitem[{Tzur {\it et~al\/}(2009)Tzur, Kafri, LeBleu, Lahav \&
  Kirschner}]{tzur2009cell}
Tzur A, Kafri R, LeBleu V, Lahav G, Kirschner M (2009) Cell growth and size
  homeostasis in proliferating animal cells. {\it Sciences STKE\/} {\bf 325}:
  167

\bibitem[{Von~Meyenburg(1969)}]{Kaspar_1969}
Von~Meyenburg HK (1969) Energetics of the budding cycle of \emph{Saccharomyces
  cerevisiae} during glucose limited aerobic growth. {\it Archives of
  Microbiology\/} {\bf 66}: 289--303

\bibitem[{Wilson(1900)}]{wilson1900cell}
Wilson E (1900) {\it The cell in development and inheritance\/}.
\newblock  Macmillan

\bibitem[{Wyart {\it et~al\/}(2010)Wyart, Botstein \&
  Wingreen}]{wyart_evaluating_2010}
Wyart M, Botstein D, Wingreen NS (2010) Evaluating Gene Expression Dynamics
  Using Pairwise {RNA} {FISH} Data. {\it PLoS Comput Biol\/} {\bf 6}: e1000979

\end{thebibliography}
\cleardoublepage

\section*{Figure 1}
\begin{figure}[!ht]
 \begin{center}
  	\includegraphics[width=0.48\textwidth ]{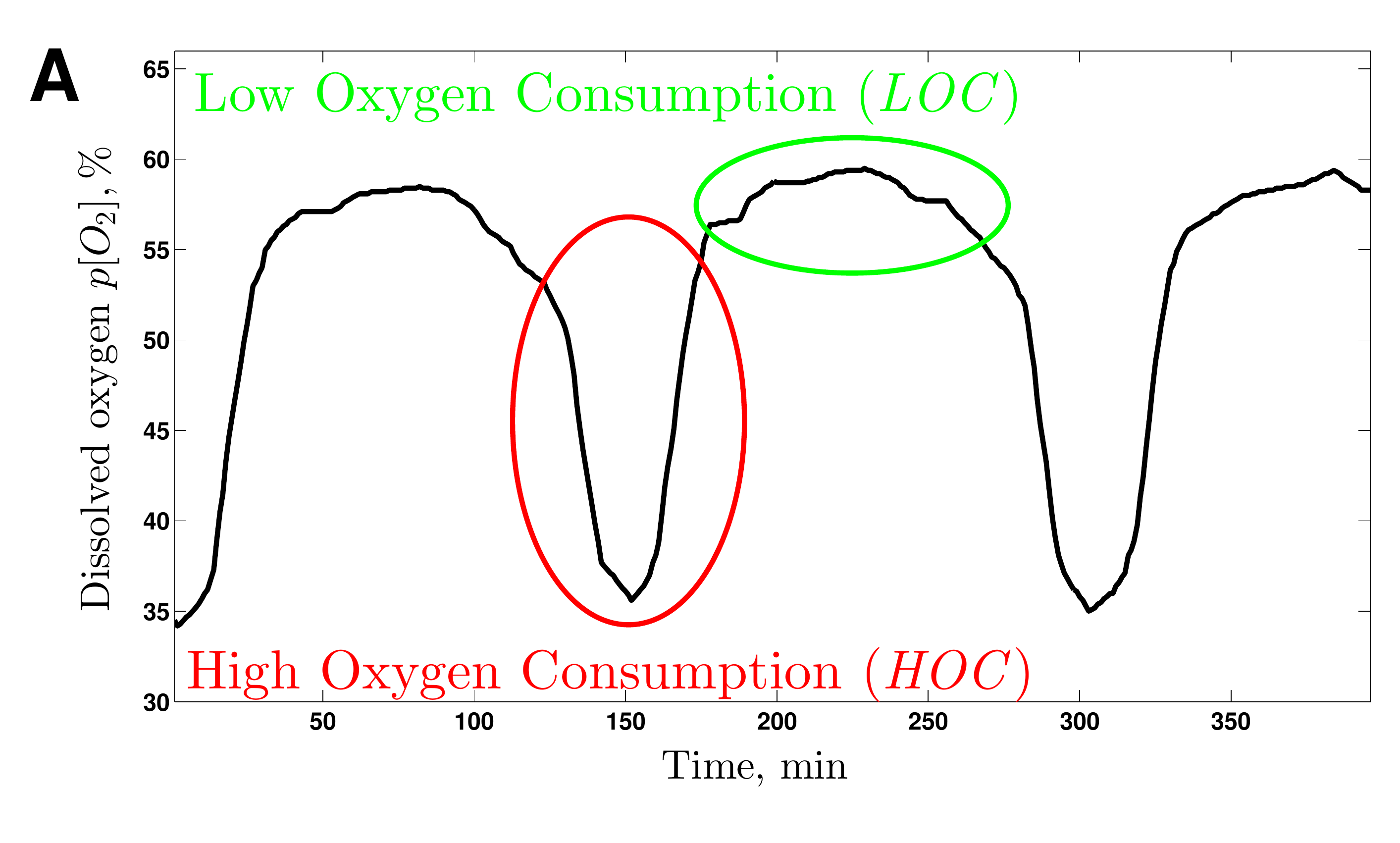}
	\includegraphics[width=0.48\textwidth ]{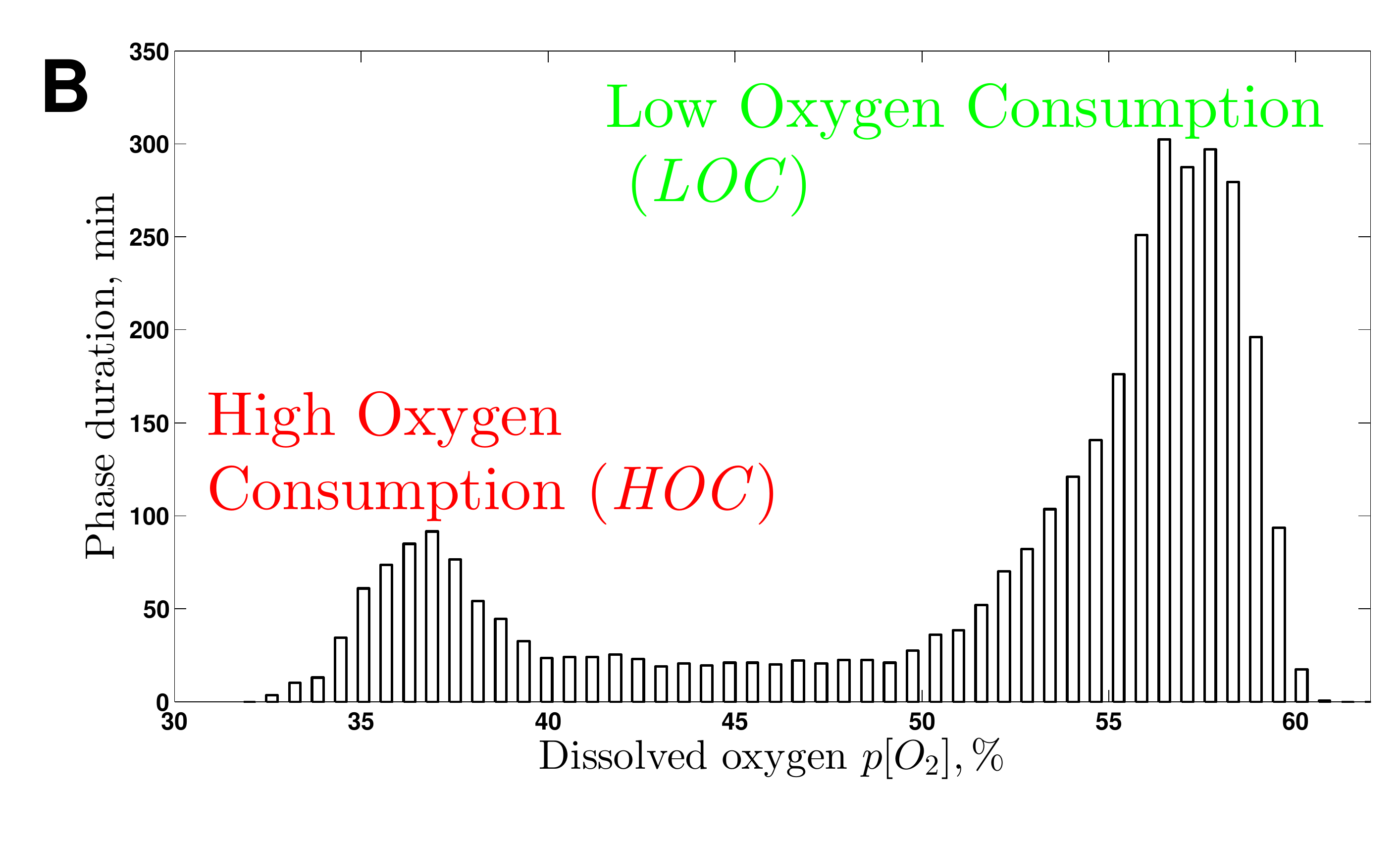} \\
    \includegraphics[width=0.78\textwidth]{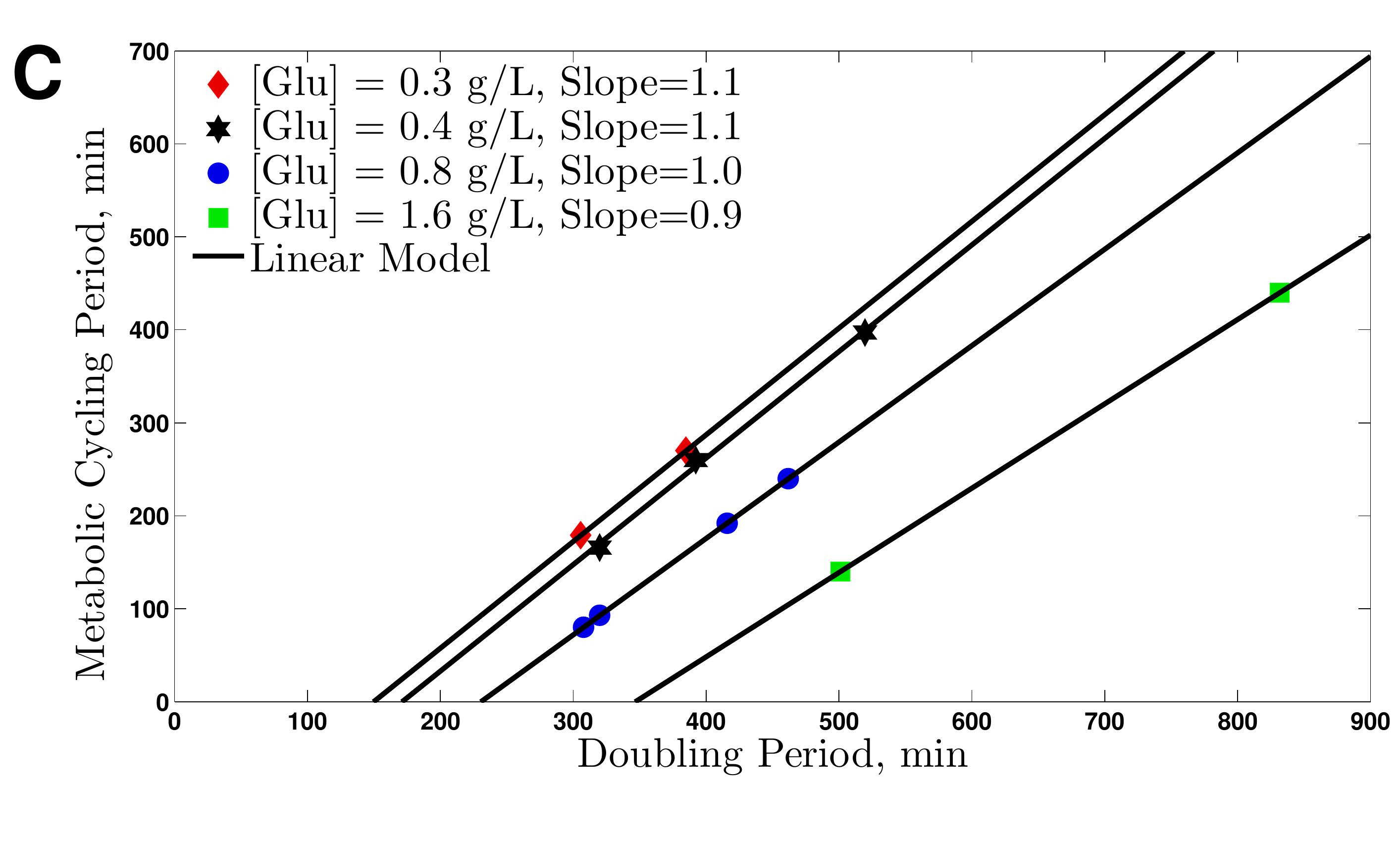} 
 \end{center}	
\end{figure}
\newpage

\section*{Figure 2}
\begin{figure}[!ht]
 \begin{center}
  	\includegraphics[width=0.98\textwidth ]{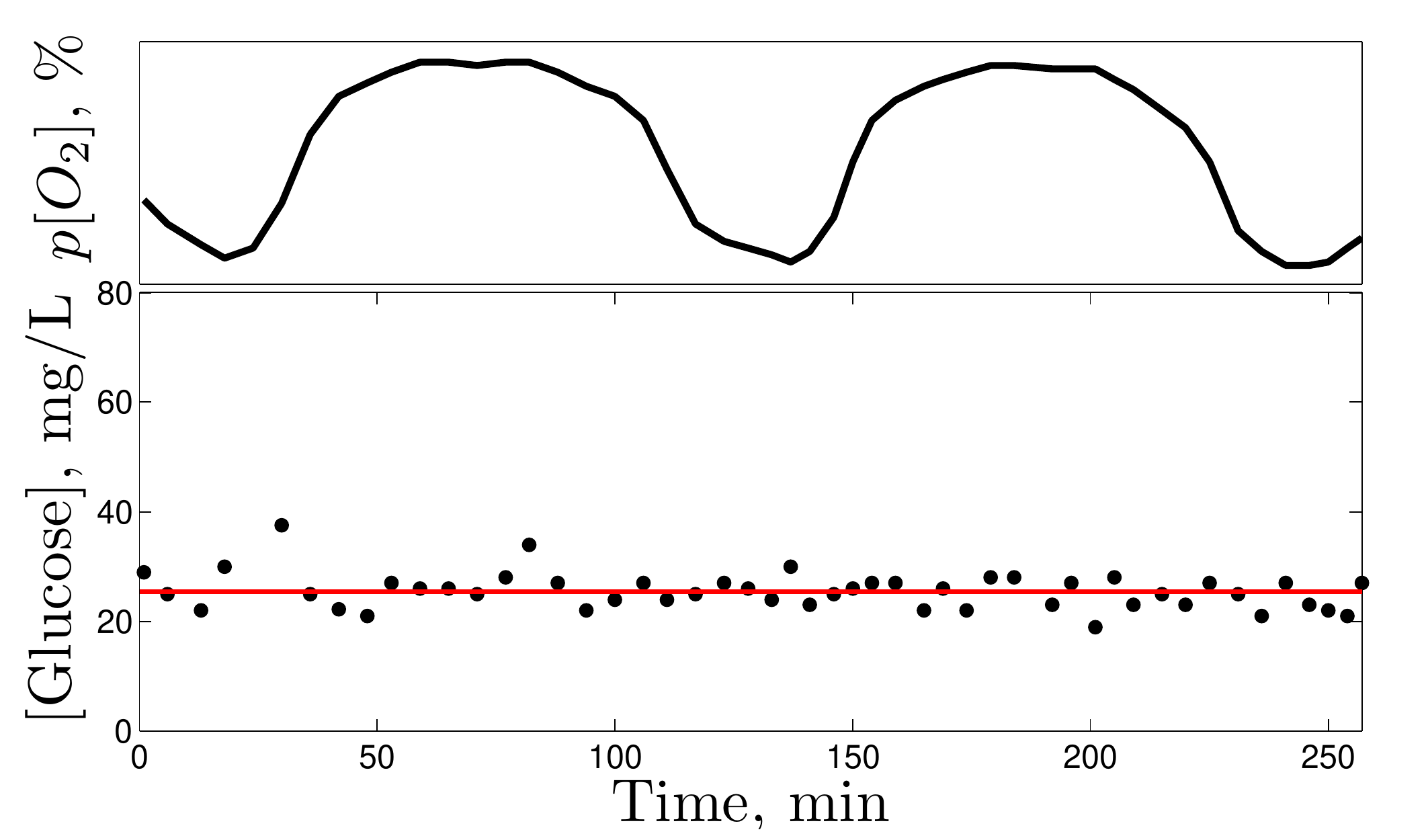}
 \end{center}
\end{figure}
\newpage

\section*{Figure 3}
\begin{figure}[!ht]
 \begin{center}
  	\includegraphics[width=0.98\textwidth ]{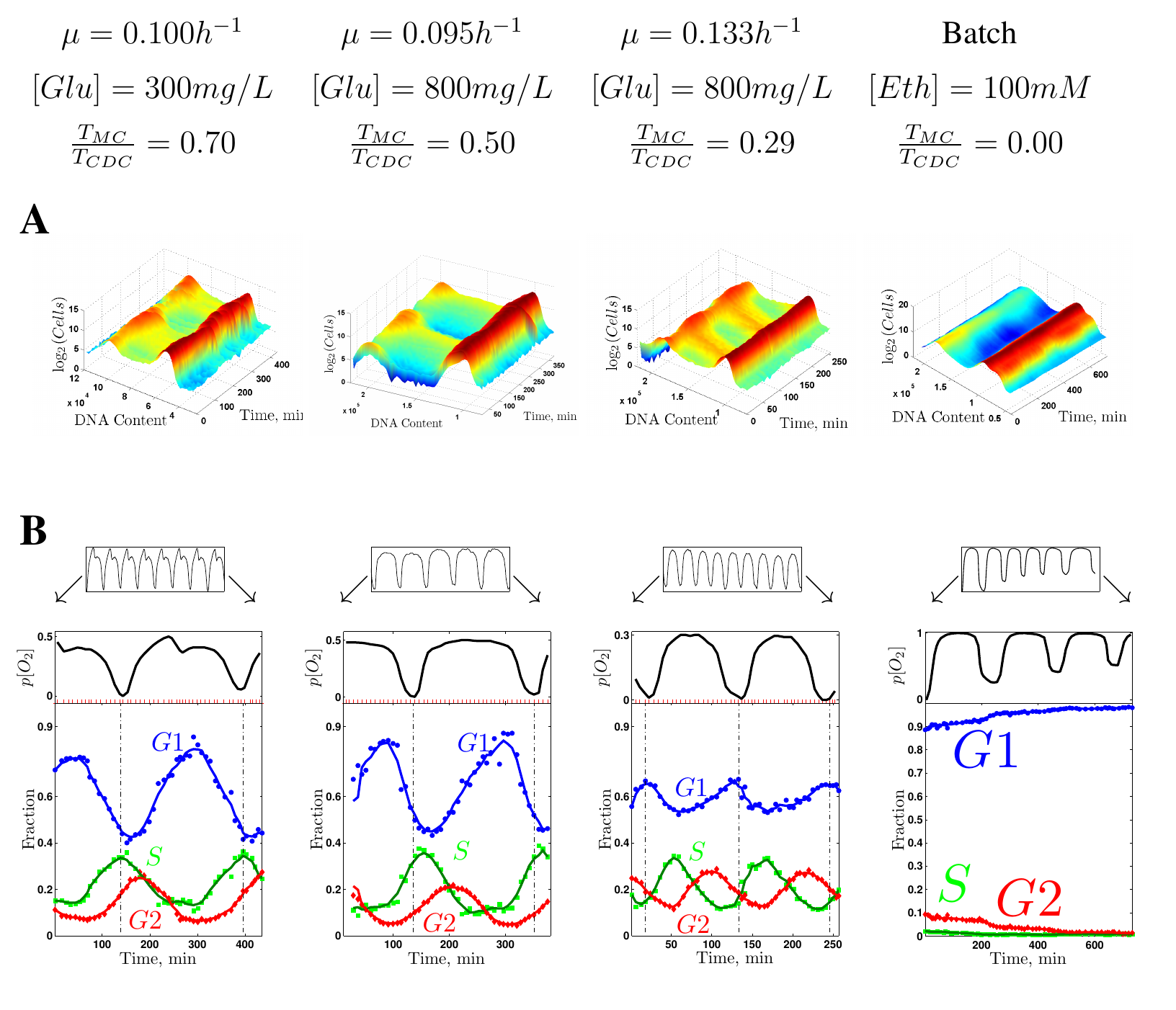}
 \end{center}
\end{figure}
\newpage

\section*{Figure 4}
\begin{figure}[!ht]
	\begin{center}
	\includegraphics[width=0.93\textwidth]{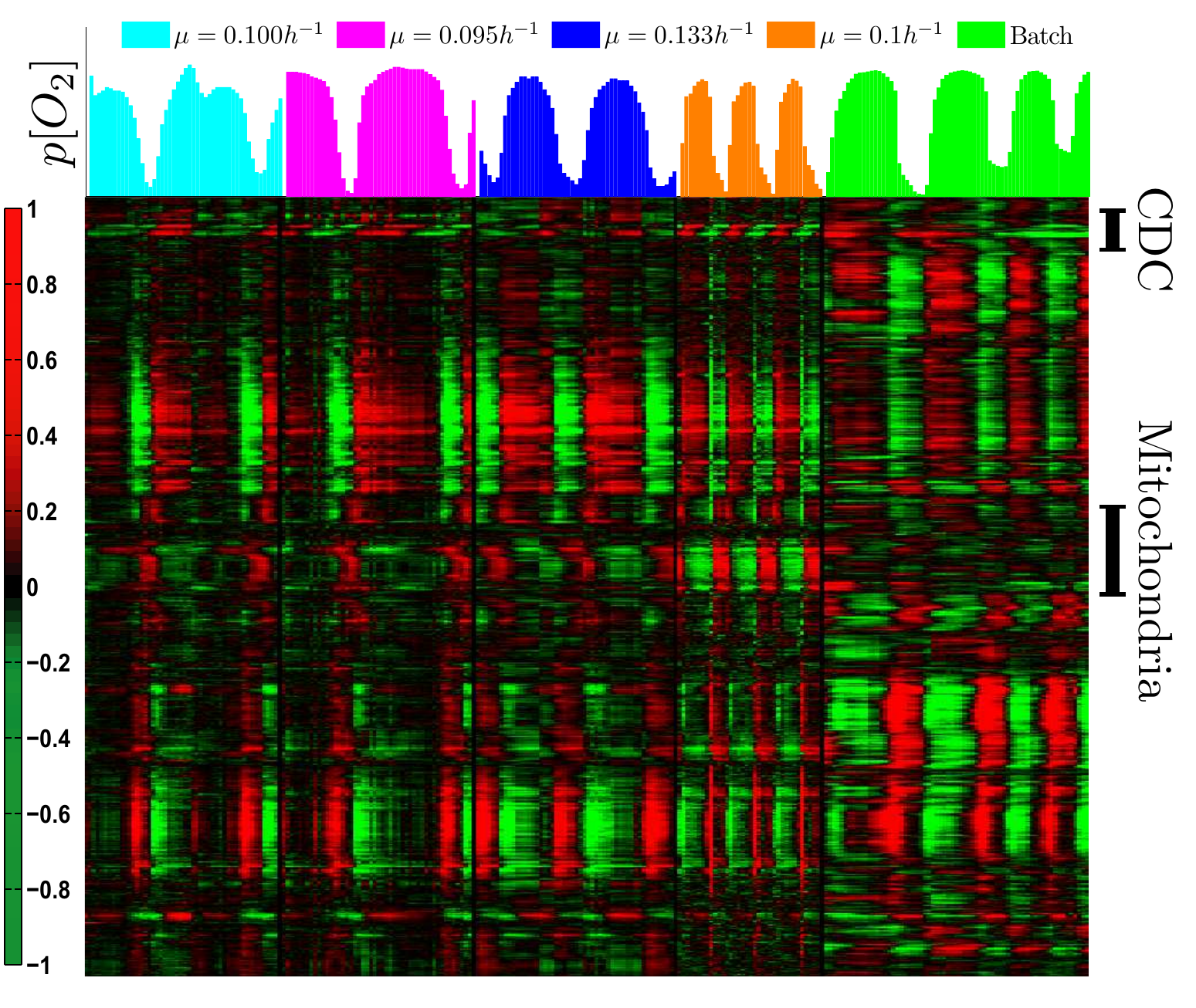}
	\end{center}
\end{figure}
\newpage

\section*{Figure 5}
\begin{figure}[!ht]
 \begin{tabular}{ p{0.01\textwidth} p{0.99\textwidth} }
  {\Large \bf A} & \vspace{0pt}
	\includegraphics[width=0.95\textwidth]{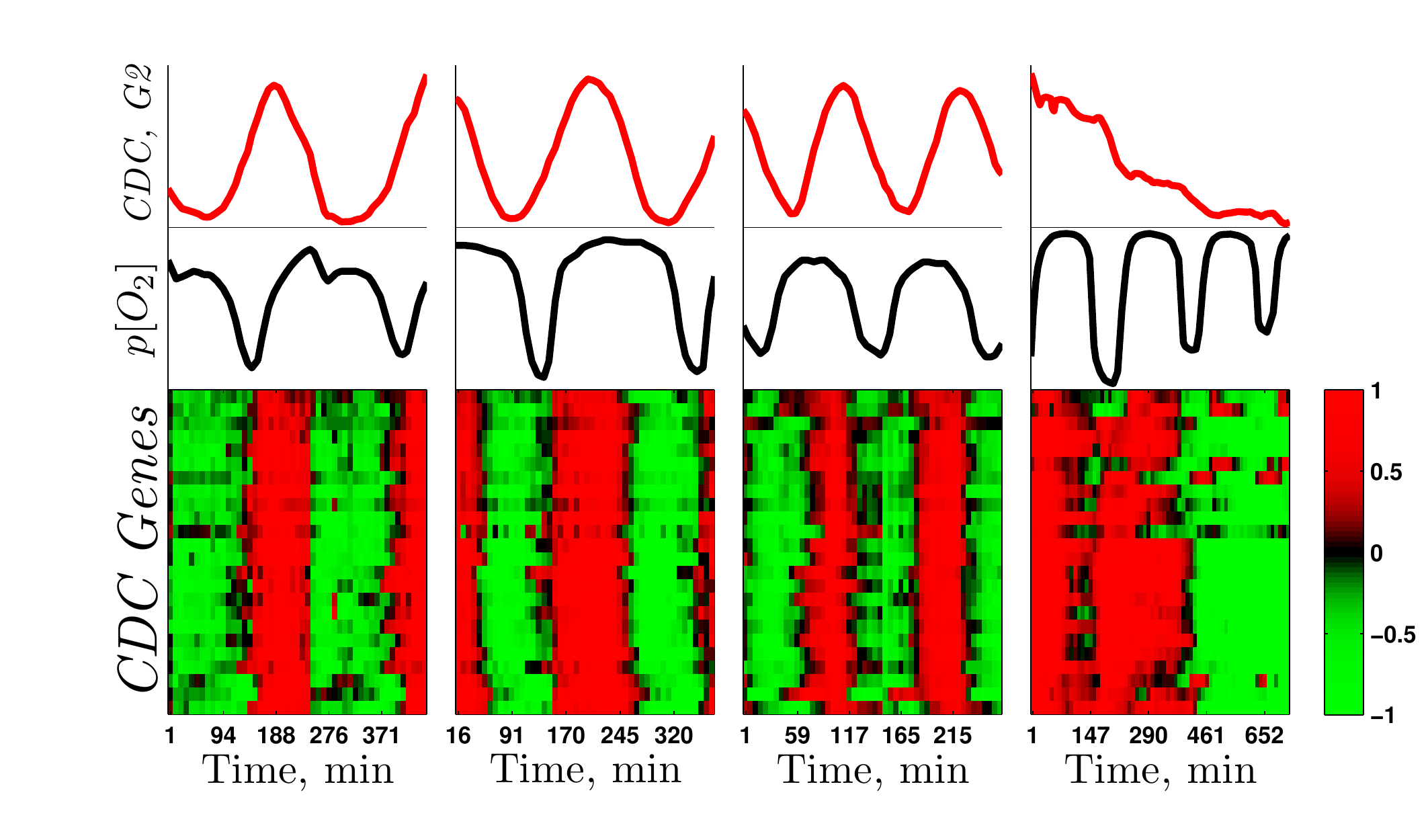}\\
  {\Large \bf B} & \vspace{0pt}
	\includegraphics[width=0.95\textwidth]{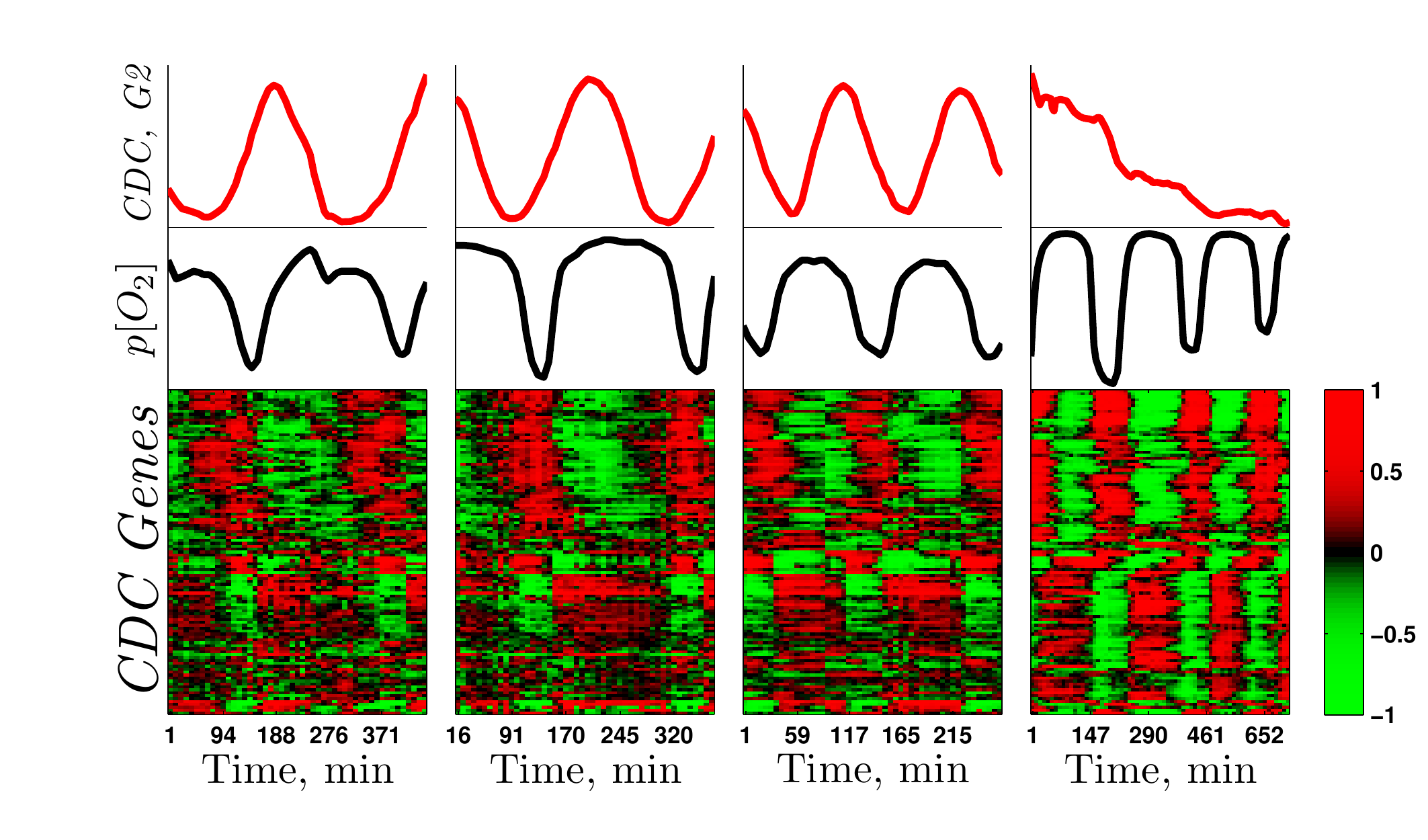}
 \end{tabular} 	
\end{figure}
\newpage

\section*{Figure 6}
\begin{figure}[!ht]
	\begin{center}
	 \includegraphics[width=0.98\textwidth]{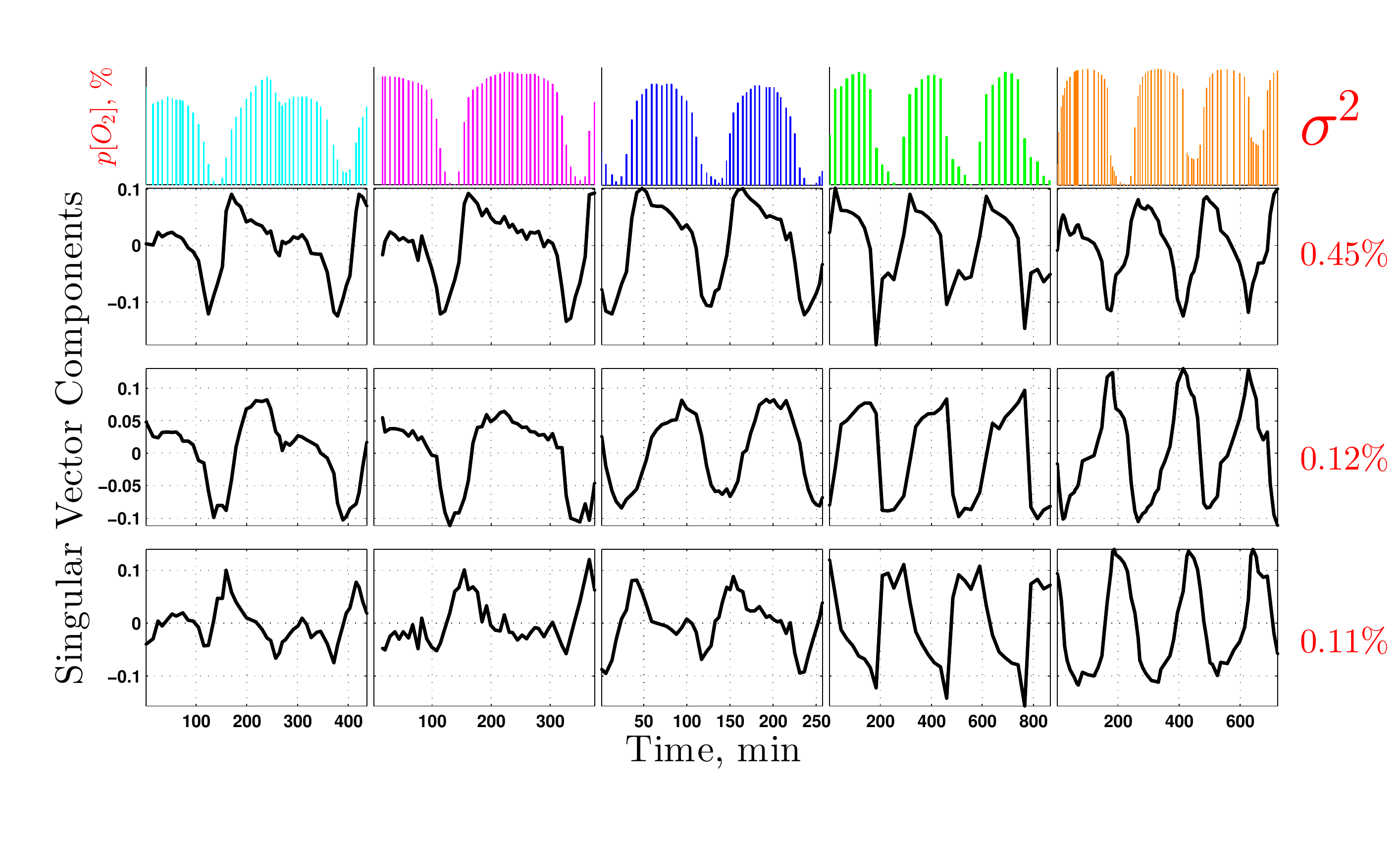}
	\end{center}	
\end{figure}
\newpage

\section*{Figure 7}
\begin{figure}[!ht]
 \begin{center}
	\includegraphics[width=0.99\textwidth]{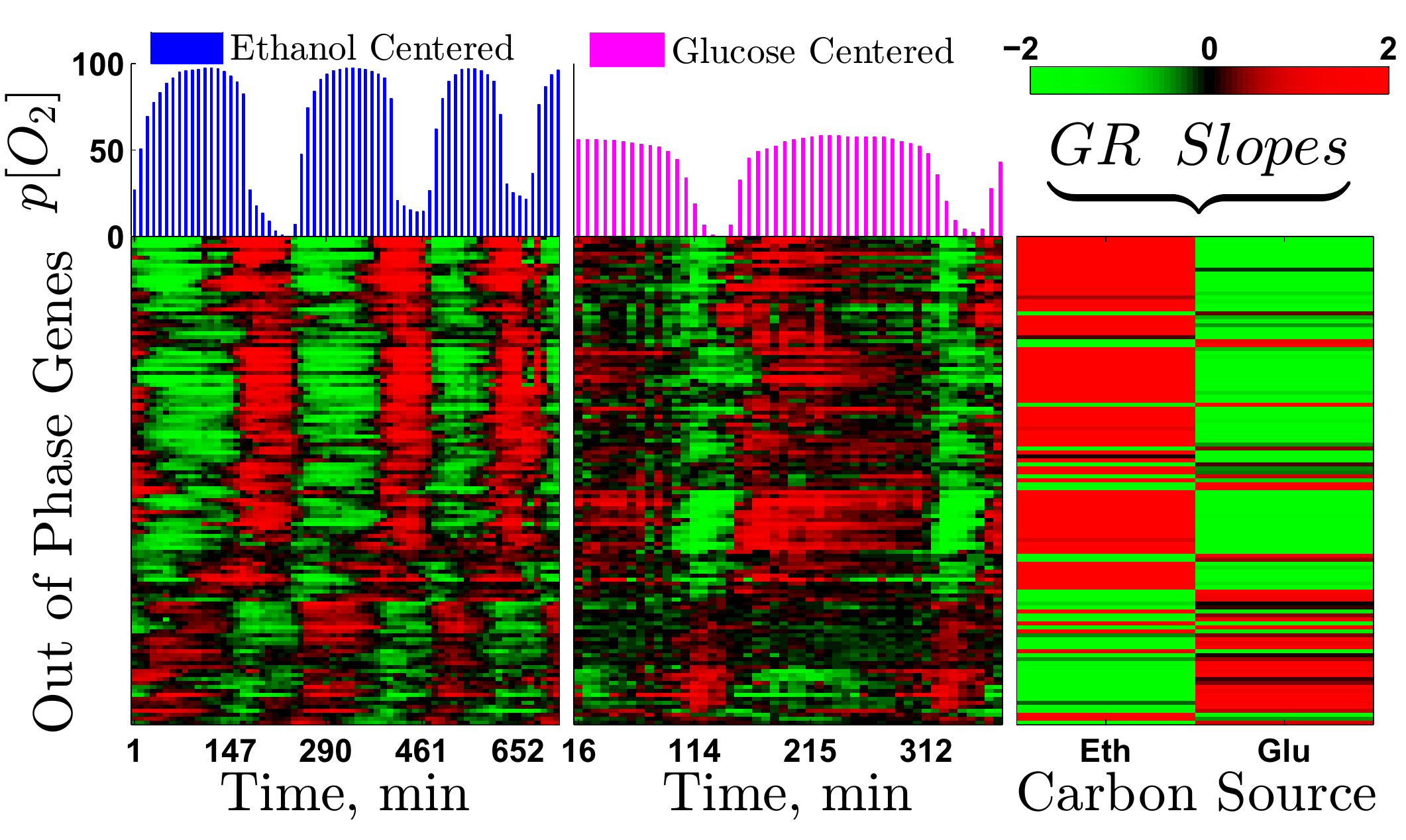}
 \end{center}
\end{figure}
\newpage

\end{spacing}
\end{document}